\documentclass[twocolumn,trackchanges]{aastex7} 
\usepackage{booktabs}
\usepackage{graphicx}   
\usepackage{tabularx}
\hypersetup{linkcolor=blue,citecolor=blue,filecolor=cyan,urlcolor=magenta}

\begin{document}

\title{The Stellar “Snake”-III: Co-evolution of Stars and Molecular Clouds Unveiled by Gaia, MWISP, and LAMOST}

\author[]{Jia-Peng, Li}
\affiliation{School of sciences, HangZhou Dianzi University, HangZhou 310018, China} 
\affiliation{Zhejiang Branch of National Astronomical Data Center, Hangzhou 310018, China}
\email{jiapeng-li@hdu.edu.cn}

\author[]{Hai-Jun, Tian} 
\affiliation{School of sciences, HangZhou Dianzi University, HangZhou 310018, China} 
\affiliation{Zhejiang Branch of National Astronomical Data Center, Hangzhou 310018, China}
\email[show]{hjtian@hdu.edu.cn}

\author[]{Chen, Wang} 
\affiliation{Institute of Astronomy and Information, Dali University, Dali 671003, China}
\email[]{wangchen@pmo.ac.cn}

\author[]{Xiang-Ming, Yang} 
\affiliation{School of sciences, HangZhou Dianzi University, HangZhou 310018, China}
\affiliation{Zhejiang Branch of National Astronomical Data Center, Hangzhou 310018, China}
\email[]{xmyang@hdu.edu.cn}

\author[]{Fan, Wang} 
\affiliation{School of Physics and Astronomy, China West Normal University, Nanchong 637002, China}
\email[]{1481616088@qq.com}
\affiliation{Purple Mountain Observatory, Chinese Academy of Sciences, Nanjing 210023, China}

\begin{abstract}
By combining multi-band data from Gaia DR3, MWISP CO, and LAMOST DR11 LSR/MSR, we investigate the co-evolution of stars and their parent molecular cloud in a snake-like stellar structure, named Snake\,III.
Based on 5-D phase-space selection, we identified 5683 member stars (median age 7.6\,Myr) across approximately $300 \times 500 \times 175$\,pc$^3$ volume, along with 12 embedded open clusters. Then we use BEEP distances combined with $^{12}$CO velocities to clearly identify the molecular clouds associated with the stellar complex in spatial and kinematics.
The molecular cloud density increases with Galactic longitude, with older open clusters forming in cavities near higher-density regions (except ASCC\,125), while young field stars currently form preferentially in present-day high-density environments, indicating that cloud density regulates the star-formation sequence.
$^{12}$CO excitation temperature, centroid velocity, velocity dispersion and H$_\alpha$ emission reveal that early feedback first compresses cloud edges to trigger new stars, then sweeps and disperses the parent clouds.
The extremely young cluster (ASCC\,125, 4.4\,Myr) lies near the densest region yet is surrounded by a shell with bidirectional density-velocity perturbations, consistent with a delayed-triggering scenario under the combined influence of UBC\,178 stellar-wind feedback and a suspected supernova blast. 
Our results naturally demonstrate that snake-like stellar structures are filamentary relics of hierarchical star formation within giant molecular clouds. 
They provide direct observational evidence that cloud density and early feedback jointly modulate the progression of star formation, offering a clear and young laboratory for studying star–cloud co-evolution.
\end{abstract}

\keywords{\uat{Young stellar objects}{1834} --- \uat{Premain sequence stars}{1290} --- \uat{Stellar populations}{1622} --- \uat{Stellar kinematics}{1608} --- \uat{Open star clusters}{1160} --- \uat{Interstellar medium}{847} --- \uat{Molecular clouds}{1072} --- \uat{Star formation}{1533} --- \uat{Stellar feedback}{1602}}

\section{Introduction}\label{sec:1}

The advent of high-precision astrometric data from the \textit{Gaia} mission has revolutionized our understanding of Galactic structure, revealing complex stellar configurations that challenge traditional models of star formation and cluster evolution. Among these discoveries, the ``Stellar Snake''—a massive hierarchical stellar filament spanning over 200\,pc—is particularly noteworthy. Initially identified by \citet{Tian2020} using \textit{Gaia} DR2 at a distance of $\sim$310\,pc, its young age (30--40\,Myr) suggests a primordial origin within Galactic giant molecular clouds \citep[GMCs;][]{Schneider1979,Su2015,Zucker2018,Soler2021,Pang2022ApJ,WangF2025AJ}, rather than formation through tidal stripping. Subsequent analysis based on \textit{Gaia} EDR3 confirmed it as a single 34\,Myr population with solar metallicity, exhibiting global outward expansion and kinematic/chemical homogeneity among its embedded clusters \citep[][hereafter W22]{Wang2022}. Studies of its mass function further revealed systematic slope variations indicative of sequential star formation \citep{Yang2024}. While these findings establish the Stellar Snake as a critical laboratory for clustered star formation, its relatively old age implies that the natal gas has fully dissipated, precluding direct investigation of star--gas co-evolution.

Similar structures have been reported contemporaneously. \citet{Kounkel2019,Kounkel2020} identified numerous Galactic filaments termed ``Stellar Strings'', proposed as stellar populations born from the same molecular reservoir. However, their physical reality has been questioned: \textit{Gaia} EDR3 analysis reveals that some candidates lack the tight spatial and kinematic coherence expected for coeval groups, exhibiting large velocity dispersions ($\sigma_{V_r}\approx 15$\,km\,s$^{-1}$) inconsistent with bound configurations and chemical homogeneity indistinguishable from field stars \citep{Zucker2022}. This ongoing debate highlights the necessity of rigorous multi-wavelength validation—combining precise kinematics, chemistry, and evidence of associated natal material—to distinguish genuine physical structures from chance alignments.

In this context, young ($<$ 10\,Myr) hierarchical structures that remain physically associated with molecular clouds are critically important for filling the aforementioned gap: they can both verify the primordial nature of the structure and directly probe star-cloud interactions. Based on this motivation, we adopt a broad definition of a “Stellar Snake” as a class of hierarchically linked young star clusters with extended structure, and have initiated a systematic census of such stellar snake-like candidates in the solar neighborhood, now extended to 3\,kpc. A full-sky catalog of candidate members with comprehensive analysis will be presented in a forthcoming work (Yang et al., in prep.). Utilizing the high-precision astrometry from \textit{Gaia} DR3, this catalog has identified hundreds of candidate structures with coherent spatial-kinematic properties.

The Stellar “Snake\,III” system studied in this paper is precisely such a unique case drawn from this census. With an age of $<$ 10\,Myr and member stars showing high consistency with residual molecular clouds in both spatial distribution and kinematics, it provides an ideal natural laboratory for investigating in-situ star formation and early feedback effects.

In this work, we present a comprehensive multi-wavelength analysis of Snake\,III. 
We combine the full astrometric power of \textit{Gaia}\,DR3 \citep{Vallenari2023} with molecular line data from the Milky Way Imaging Scroll Painting (MWISP) project \citep{Yang2025arXiv251208260Y} and complementary spectroscopy from surveys including LAMOST \citep{Cui2012} and APOGEE \citep{Majewski2017}. 
We employ an FoF algorithm to identify the stellar members of Snake\,III and characterize its spatial and kinematic structure. 
Furthermore, we analyze the $\mathrm{^{12}CO}$ emission from the associated cloud complex. 
Under the assumption of local thermodynamic equilibrium \citep[LTE;][]{Pineda2008,Pineda2010}, we use these lines to derive the physical properties of the gas, mapping its distribution, mass, density, and kinematics to quantify the interplay between the stellar population and its birth cloud environment.

This paper is structured as follows. 
Section~\ref{sec:2} details the data sources and methodologies for processing stellar and molecular line observations. 
Section~\ref{sec:3} presents the integrated results, combining the spatial-kinematic properties of the stellar population with the morphology and physical state of the molecular cloud. 
We discuss the implications of these findings in Section~\ref{sec:4}, focusing on the co-evolutionary history of the system and assessing the role of stellar feedback in regulating star formation. 
Our conclusions are summarized in Section~\ref{sec:5}.

Throughout the paper, we adopted the solar motion $(U_\odot, V_\odot, W_\odot) = (9.58, 10.52, 7.01)$~km~s$^{-1}$\citep{Tian2015} with respect to the local standard of rest (LSR), and the solar to Galactic center radius and vertical height $(R_0, Z_0) = (8.27, 0.0)$~kpc\citep{Schonrich2012}.
In the coordinate system of the gnomonic projection, we use \textit{ l$^*$} to denote Galactic longitude ($\mu_{l^*}$ = $\mu_l$ cos $b$), and obtain distance by a direct parallax inversion with $d=1000.0\,\omega^{-1}$ (pc).

\section{Data and Analysis}\label{sec:2}

In this section, we describe the data sources and reduction procedures for Snake\,III and the associated molecular clouds.

\subsection{Stellar Snake\,III}\label{subsec:2.1}

To investigate the co-evolution of stars and molecular clouds, we searched for specific targets from the all-sky “Stellar Snake” census catalog that satisfy two critical criteria: extremely young age ($<$ 10\,Myr) and coverage by high-resolution molecular cloud data. 
Since the MWISP project provides high-quality CO observations specifically within the Galactic plane ($-5^\circ < b < 5^\circ$), we strategically focused our selection on this region. 
This process identified “Snake\,III” as the ideal laboratory. 
It is a prominent candidate stretching almost $25^\circ$ in Galactic longitude ($90^\circ < l < 115^\circ$) at low Galactic latitude ($0^\circ < b < 5^\circ$), excellently covered by matching available gas data while retaining a rich population of member stars. 
Therefore, in this section, we refine the member selection specifically for Snake\,III to ensure high fidelity for subsequent detailed analysis.

\subsubsection{Selection of Snake\,III Member Candidates}\label{subsubsec:2.1.1}

As in our previous serial works \citep[W22;][]{Tian2020,Yang2024}, we identified an initial sample of 10,288 candidate member stars of Snake\,III by applying a FoF algorithm in the 5-D phase space (i.e., \textit{l}, \textit{b}, $\mu_{l^*}$, $\mu_b$ and distance) using the ROCKSTAR code \citep{Behroozi2013}.
After applying quality cuts ($\omega/\sigma_\omega > 10.0$; $G < 18$~mag; RUWE $<$ 1.4), we retained a final sample of 5,683 high-confidence members.

\begin{figure*}
    \centering
    \includegraphics[width=1\linewidth]{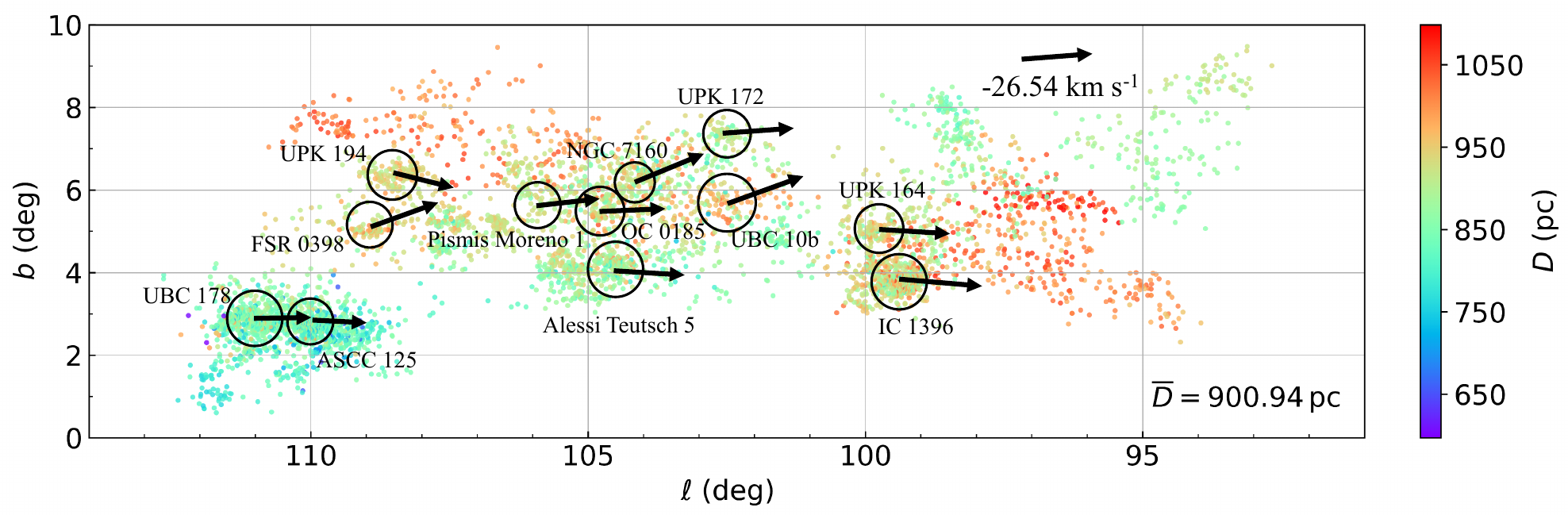}
    \caption{Spatial distribution of the member stars of Snake\,III. The color bar encodes distance. Black circles mark the centers and angular sizes of the open clusters. The black arrow in the top-right corner indicates the median tangential velocity of the entire sample, while the remaining arrows show the median tangential velocities of the individual cluster samples, with their lengths proportional to the velocity magnitudes. All velocities are given with respect to the LSR.}
    \label{fig.1}
\end{figure*}

\subsubsection{Open Clusters in Snake\,III}\label{subsubsec:2.1.2}

In order to study the co-evolution of stars and molecular clouds, it is essential to identify open clusters in Snake\,III.
Following a systematic search and investigation using an all-sky clusters catalog from \citet{Hunt2023}, we have identified 12 known open clusters in Snake\,III.
The fundamental parameters of the 12 open clusters were derived using the package Automated Stellar Cluster Analysis \citep[ASteCA;][]{Perren2015}, with the same configuration as described in \citet{Yang2024}.
Subsequently, for the initial sample within $r_{cl}$, we retained all stars whose parallax and proper-motion falls within the range of the median value $\pm$3$\sigma$.
We thus obtained the final samples of the member stars for all open clusters, with the corresponding counts $N$ and the fundamental parameters listed in Table \ref{tabel.1}.

\begin{table*}
\centering
\caption{Derived parameters of Stellar Snake\,III Open Clusters} \label{tabel.1} 
\resizebox{\textwidth}{!}{
\begin{tabular}{cccccccccccc}
\toprule\toprule
Clusters & $N$ & $l$ & $b$ & $r_{\text{cl}}$ & $\bar{V}_l$ & $\bar{V}_b$ & \multicolumn{2}{c}{$\bar{D}$} & \multicolumn{3}{c}{log$\,$Age} \\
\midrule
&&& & &&& ASteCA& Hunt's catalog& Sagitta($pms>$ 0.1)& ASteCA& Hunt's catalog\\
\midrule
&& \multicolumn{2}{c}{deg} & arcmin & \multicolumn{2}{c}{km/s} & \multicolumn{2}{c}{pc} & \multicolumn{3}{c}{dex} \\
\midrule
Alessi Teutsch$\,$5 & 123 & 104.48360 & 4.21961 & 30.33$^{+0.43}_{-3.86}$ & -26.35 & -0.43 & 895.44$\pm2.38$ & 875.52$\pm1.78$ & 6.97$\pm0.04$ & 7.03$\pm0.05$ & 6.81$^{+0.16}_{-0.17}$ \\
\midrule
ASCC$\,$125 & 332 & 110.16833 & 2.70908 & 21.36$^{+2.98}_{-1.49}$ & -20.18 & -0.25 & 845.32$\pm1.66$ & 828.18$\pm2.02$ & 6.67$\pm0.01$ & 6.61$\pm0.05$ & 6.56$^{+0.08}_{-0.15}$ \\
\midrule
FSR$\,$0398 & 59 & 108.88551 & 5.03454 & 23.61$^{+0.28}_{-11.51}$ & -27.33 & 8.78 & 944.52$\pm2.57$ & 924.58$\pm1.48$ & 7.01$\pm0.05$ & 7.39$\pm0.04$ & 7.15$^{+0.20}_{-0.19}$ \\
\midrule
IC$\,$1396 & 297 & 99.30483 & 3.72840 & 30.37$^{+5.23}_{-7.07}$ & -32.26 & -1.95 & 924.00$\pm1.81$ & 905.69$\pm1.19$ & 6.73$\pm0.02$ & 6.60$\pm0.07$ & 6.55$^{+0.07}_{-0.13}$ \\
\midrule
NGC$\,$7160 & 70 & 104.00830 & 6.44018 & 18.78$^{+0.60}_{-4.15}$ & -27.27 & 9.77 & 926.00$\pm2.66$ & 894.88$\pm1.45$ & 7.09$\pm0.01$ & 7.13$\pm0.06$ & 6.95$^{+0.16}_{-0.17}$ \\
\midrule
OC$\,$0185 & 53 & 104.83637 & 5.39743 & 24.79$^{+7.98}_{-1.43}$ & -24.25 & 0.98 & 926.26$\pm2.90$ & 908.82$\pm1.07$ & 7.05$\pm0.02$ & 6.68$\pm0.12$ & 6.81$^{+0.13}_{-0.17}$ \\
\midrule
Pismis Moreno$\,$1 & 40 & 106.69290 & 5.29580 & 21.44$^{+8.23}_{-8.84}$ & -25.19 & 2.76 & 921.57$\pm1.49$ & 909.09$\pm1.91$ & 6.90$\pm0.05$ & 7.86$\pm0.07$ & 6.74$^{+0.12}_{-0.17}$ \\
\midrule
UBC$\,$10b & 68 & 102.65325 & 5.79315 & 35.73$^{+0.20}_{-15.75}$ & -29.35 & 9.74 & 955.98$\pm3.87$ & 937.55$\pm1.76$ & 7.02$\pm0.02$ & 7.41$\pm0.09$ & 7.75$^{+0.23}_{-0.20}$ \\
\midrule
UBC$\,$178 & 285 & 110.95392 & 2.75898 & 29.73$^{+4.73}_{-12.94}$ & -20.99 & 0.78 & 850.46$\pm1.50$ & 835.06$\pm1.79$ & 6.76$\pm0.01$ & 7.08$\pm0.05$ & 6.58$^{+0.08}_{-0.15}$ \\
\midrule
UPK$\,$164 & 63 & 99.86700 & 4.97494 & 25.13$^{+7.75}_{-10.67}$ & -26.31 & -0.24 & 932.47$\pm2.00$ & 908.78$\pm1.81$ & 6.91$\pm0.04$ & 7.04$\pm0.14$ & 6.60$^{+0.1}_{-0.12}$ \\
\midrule
UPK$\,$172 & 46 & 102.55905 & 7.318778 & 23.02$^{+5.17}_{-8.75}$ & -27.13 & 1.70 & 913.35$\pm2.21$ & 899.71$\pm1.54$ & 7.09$\pm0.01$ & 7.41$\pm0.14$ & 7.03$^{+0.2}_{-0.19}$ \\
\midrule
UPK$\,$194 & 75 & 108.17579 & 6.33928 & 27.28$^{+8.23}_{-4.97}$ & -23.16 & -3.52 & 941.30$\pm2.20$ & 914.09$\pm1.58$ & 6.99$\pm0.08$ & 7.42$\pm0.22$ & 7.55$^{+0.23}_{-0.25}$ \\
\bottomrule
\end{tabular}
}
\end{table*}

\subsubsection{Spatial-Kinematics Distribution}\label{subsubsec:2.1.3}

Having assembled 5683 candidate member stars of Snake\,III and the fundamental parameters together with member stars of 12 open clusters, we can now characterize the whole Stellar Snake\,III through a detailed spatial-kinematic distribution.
This will provide a solid foundation for our subsequent investigation of the association between stars and molecular clouds.

\begin{figure}
    \centering
    \includegraphics[width=1.05\linewidth]{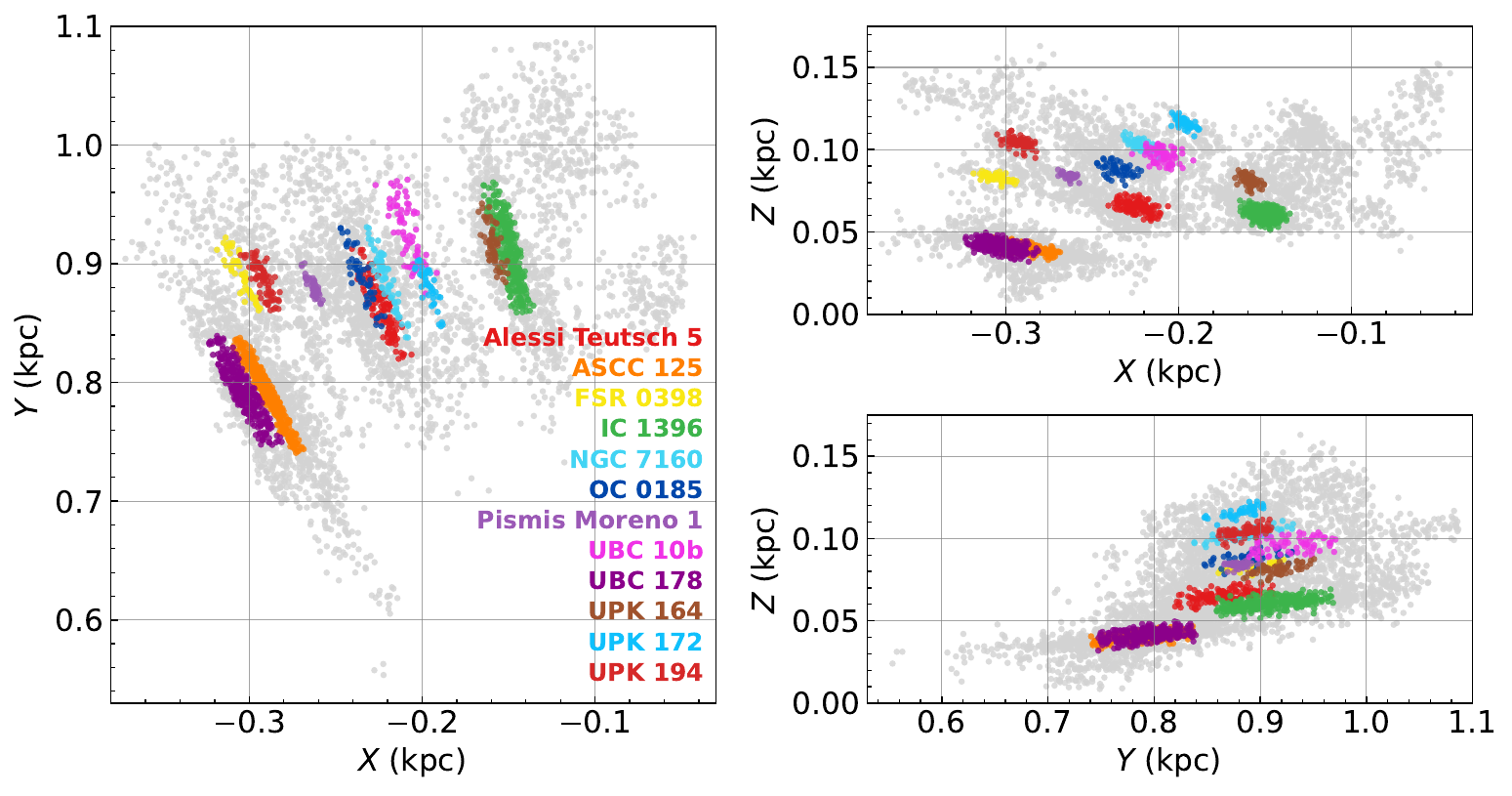}
    \caption{Spatial distribution of Snake\,III member stars (gray dots) and cluster member stars (shown as dots in various colors) in Cartesian coordinates (\textit{X, Y, Z}). The coordinates of the Sun are (\textit{X, Y, Z}) = (0.0, 0.0, 0.0)\,kpc}
    \label{fig.2}
\end{figure}

\begin{figure}
    \centering
    \includegraphics[width=0.95\linewidth]{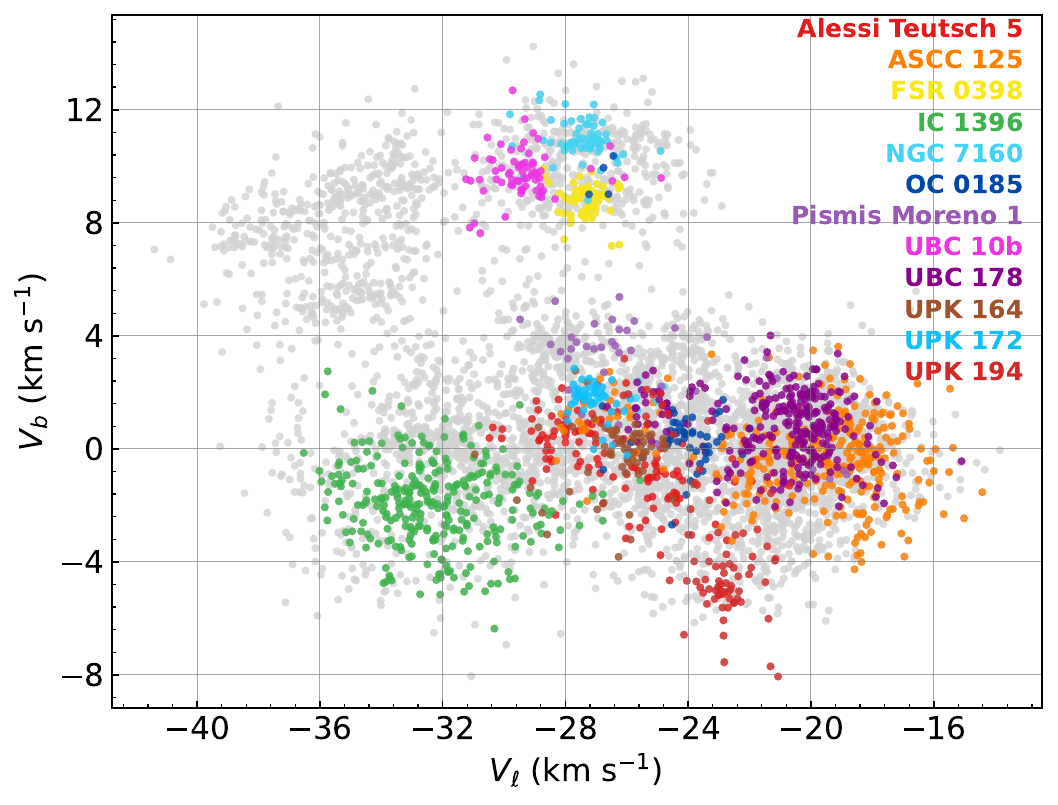}
    \caption{Tangential velocity in ($l, b$) distribution of Snake\,III member stars (gray dots) and cluster member stars (shown as dots in various colors). All velocities are given with respect to the LSR.}
    \label{fig.3}
\end{figure}

Figure \ref{fig.1} and \ref{fig.2} presents the spatial distribution of Snake\,III member stars.
The Snake\,III extends approximately 300$\times$500$\times$175\,pc, encompassing a large volume.
As shown, 12 open clusters (with different colors) are well bridged with the main structure of Snake\,III.
The open clusters can also be clearly distinguished in the tangential velocity ($V_l$--$V_b$) plane (Figure \ref{fig.3}), and we added the median tangential velocity $\bar{V}_l$ and $\bar{V}_b$ to Table \ref{tabel.1}.

\begin{figure*}
    \centering
    \includegraphics[width=1\linewidth]{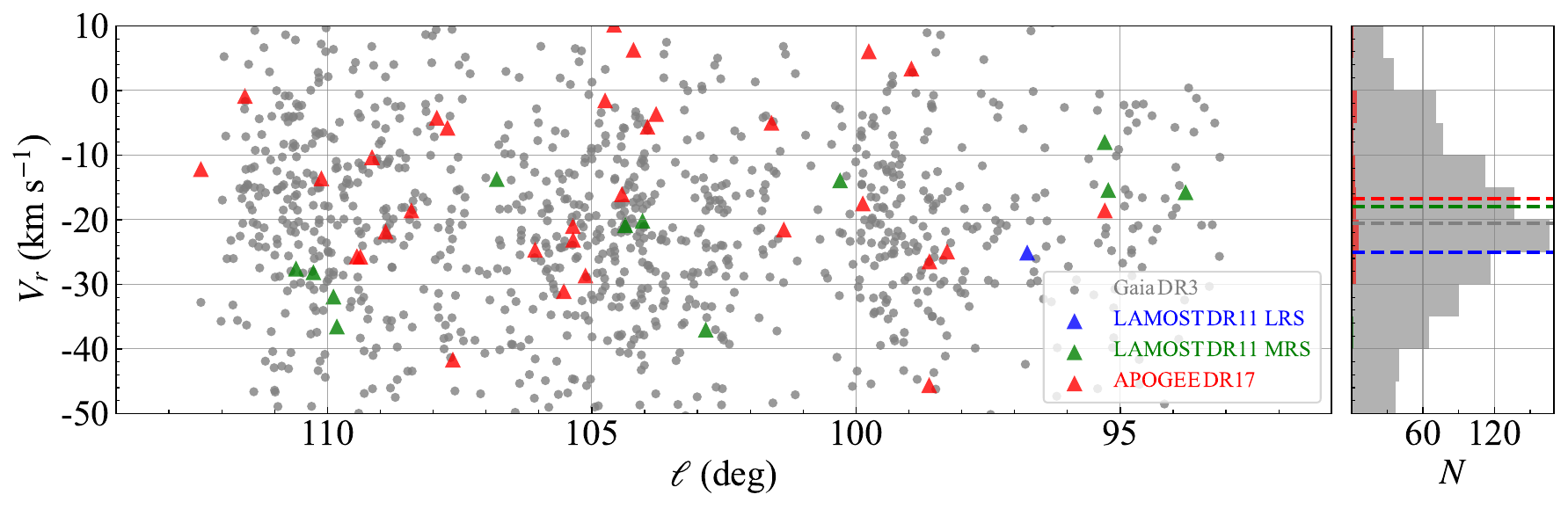}
    \caption{Distribution of the observed {$V_r$}s for all member candidates along Galactic longitude $l$. The {$V_r$}s are obtained from the multiple surveys (color-coded as specified in the legend). The histograms of {$V_r$}s are displayed in the right sub-panel. The median values are marked with dashed lines. The color-coding is the same as the main panel.}
    \label{fig.4}
\end{figure*}

Among the 5683 candidate member stars of Snake\,III, 1035 stars have {radial velocity} ({$V_r$}) measurements: 1001 from \textit{Gaia}\,DR3, 28 from APOGEE\,DR14 (3 of which also have \textit{Gaia}'s {$V_r$}), 13 from LAMOST\,DR12 MRS (4 of which also have \textit{Gaia}'s {$V_r$}) and 1 from LAMOST\,DR12 LRS (both of which also have \textit{Gaia}'s {$V_r$}). 
For stars with multiple {$V_r$} determinations from different surveys (in addition to the 3 stars with high-resolution spectra from APOGEE\,DR14), we adopt the median value as the final radial velocity.
Along the \textit{l}-direction, no significant {$V_r$} trend is observed, but the {$V_r$} distribution remains well fit by a Gaussian (Figure \ref{fig.4}).
The median {$V_r$} with uncertainty is $-20.57\pm0.51$\,km\,s$^{-1}$.

\subsubsection{Ages of Snake\,III}\label{subsubsec:2.1.4}

Current age estimates for young stellar objects (YSOs, $<$ 10\,Myr) remain uncertain, and even coeval cluster member stars appear widely scattered on the CMD.
We therefore adopt two complementary metrics to characterize the ages of Snake\,III candidate member stars: 
one approach employs a trained neural network model \citep[Sagitta;][]{McBride2021}, using \textit{Gaia}\,DR3 and \textit{2MASS} photometry plus parallaxes to systematically identify pre-main-sequence (PMS) stars among Snake\,III and to estimate their ages;
and another one treats the 12 open clusters as regional probes, deriving cluster ages via isochrone fitting. 
This allows us to map any age gradient across Snake\,III in detail.

\begin{figure}
    \centering
    \includegraphics[width=1\linewidth]{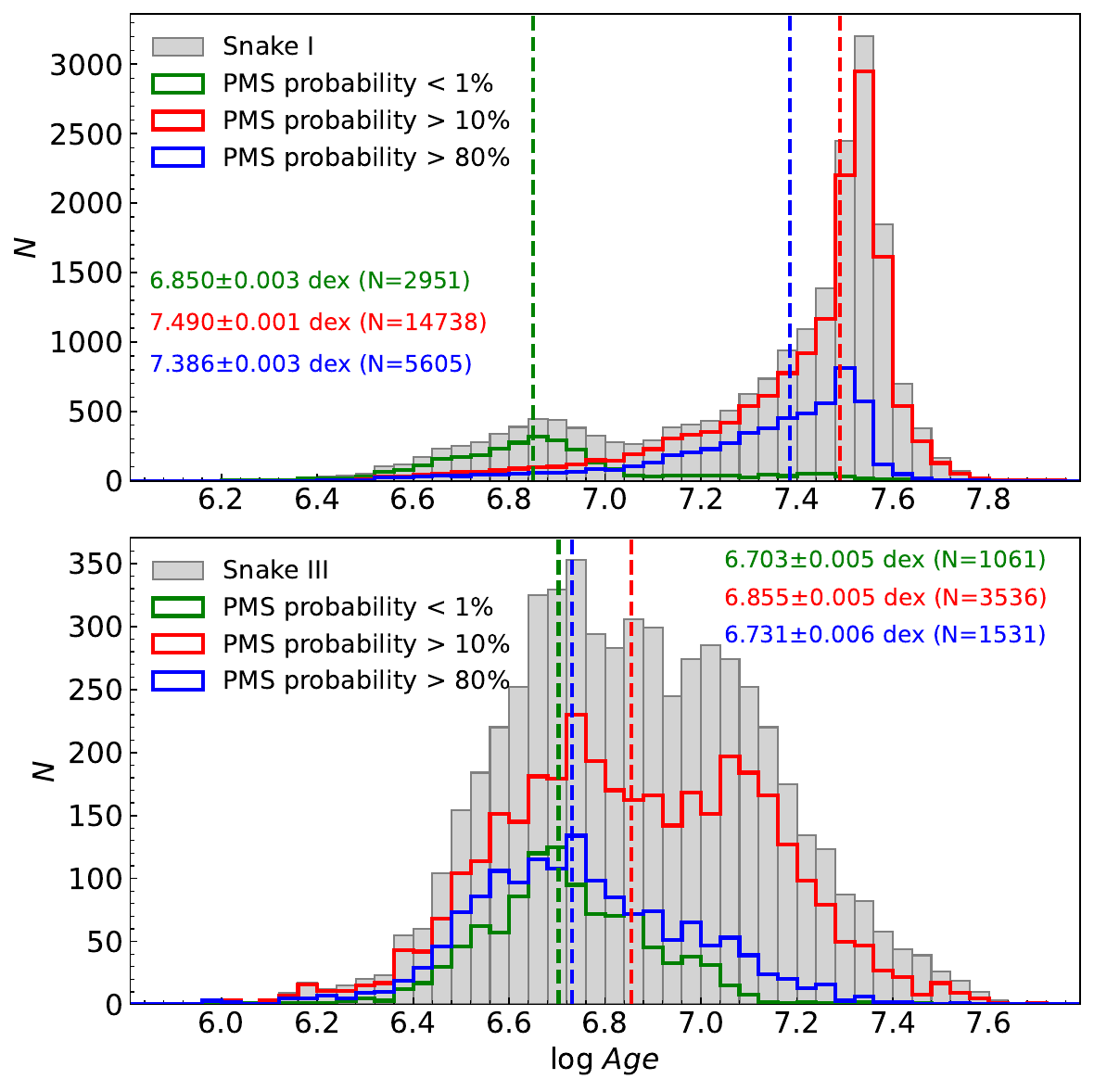}
    \caption{log\,Age distribution histograms of the Snake\,I member stars (upper panel) and Snake\,III member stars (lower panel) fitting by Sagitta. Contour lines of different colors delineate intervals of PMS probability (green for \textit{pms} $<$ 0.01, red for \textit{pms} $>$ 0.1, and blue for \textit{pms} $>$ 0.8), and the median log\,Age and number of stars calculated for each interval is marked in the corresponding color.}
    \label{fig.5}
\end{figure}

We first applied the Sagitta model to the member stars of Snake\,I (W22) for validation.
For member stars selected by Sagitta with $pms$ scores $<$ 0.01, the resulting age estimates (median log\,Age = 6.850$\pm0.003$\,dex, Figure \ref{fig.5}) are systematically lower than those reported from W22, therefore, because their extremely low $pms$ scores render the age estimates unreliable, we exclude these stars from further analysis;
for member stars with $pms$ scores $>$ 0.1, the age estimates concentrated between 30--40\,Myr, peaking at 34\,Myr (median log\,Age = 7.490$\pm0.001$\,dex), in excellent agreement with W22, so we therefore adopt age estimates as reliable only when $pms$ scores $>$ 0.1; 
for member stars with $pms$ scores $>$ 0.8, the age estimates (median log\,Age = 7.386$\pm0.003$\,dex) are even closer to the PMS stars.

As shown in the lower panel of Figure \ref{fig.5}, the younger Snake\,III sample yields a higher fraction of PMS stars from Sagitta. 
For member stars with $pms$ scores $>$ 0.1, the ages estimates between 4--15\,Myr with a median log\,Age = 6.855$\pm0.005$\,dex, and we adopt this estimate as reliable reference ages for Snake\,III member stars;
for member stars with $pms$ scores $>$ 0.8, the ages estimates between 2.5--10\,Myr (median log\,Age = 6.731$\pm0.006$\,dex), and this value serves as the age reference for stars that are almost certainly PMS stars.
We have added the age estimates for each cluster obtained by Sagitta with $pms$ $>$ 0.1 to Table \ref{tabel.1} as a comparison.

For each cluster, we refine the age fit in the CMD using ASteCA \citep{Perren2015}, which is based on the PARSEC \citep{Bressan2012,Marigo2017} isochrone models with metallicity $Z$ $=$ 0.01--0.02 to estimate the ages.
We apply it to 12 open clusters and list the derived ages in Table \ref{tabel.1}.

Because these clusters are young, their ages are difficult to determine. 
For the four clusters with more than 100 member stars (Alessi Teutsch\,5, ASCC\,125, IC\,1396 and UBC\,178), their ample membership numbers enable relatively reliable age estimates, which are also the key targets of our follow-up work. 
Therefore, we averaged their age estimates obtained from the Sagitta model (for $pms$ $>$ 0.1) and the ASteCA model as their final age estimates for further discussion:
Alessi Teutsch\,5, $10.00^{+1.09}_{-0.98}$\,Myr; ASCC\,125, $4.37^{+0.31}_{-0.30}$\,Myr; IC\,1396, $4.62^{+0.51}_{-0.45}$\,Myr; UBC\,178, $8.32^{+0.59}_{-0.56}$\,Myr.

Table \ref{tabel.1} also lists age estimates of the 12 open clusters from the catalog released by \citet{Hunt2023} for comparison.

\subsection{CO Molecular Clouds}\label{subsec:2.2}

CO and its isotopologues are key tracers of molecular clouds \citep[MCs;][]{Dame2001}. 
Under the assumption of LTE \citep{Pineda2008,Pineda2010}, these lines yield a host of physical parameters. 
$^{12}$CO is generally optically thick \citep{Goldsmith1999}; its excitation temperature can be derived \citep{Bourke1997} and, via the $X$-factor, the column density of H$_2$ is obtained from the observed CO intensity \citep[the empirical CO-to-H$_2$ conversion factor $X=N(\text{H}_2)/I_{\text{CO}}$;][]{Dame2001,Bolatto2013}. 
$^{13}$CO and C$^{18}$O are typically thinner \citep{Hacar2016}, allowing N($^{13}$CO) and N(C$^{18}$O) to be calculated directly under LTE. 
Far-ultraviolet (FUV: 6--13.6\,eV) photons from massive stars regulate the structure, dynamics, and chemical evolution of MCs \citep{Hollenbach1997,Wang2019}. 

\subsubsection{MWISP CO Map}\label{subsubsec:2.2.1}

We utilized the $^{12}$CO, $^{13}$CO and C$^{18}$O in the J $=$ 1--0 data from the Milky Way Imaging Scroll Painting (MWISP) project \citep{Yang2008}.
Observations from the MWISP project cover a total area of approximately 2,300\,deg$^2$ near the northern Galactic plane, spanning Galactic longitude $10^\circ-230^\circ$ and Galactic latitude $\pm5^\circ$ conducted with the PMO 13.7\,m telescope at Delingha \citep{Yang2025arXiv251208260Y}. 
The half-power beamwidth of the telescope was approximately 49$^{\prime\prime}$ for $^{12}$CO and 51$^{\prime\prime}$ for $^{13}$CO and C$^{18}$O, with a pointing accuracy better than 4$^{\prime\prime}$ for each observing epoch and a velocity resolution of 0.16--0.17\,km~s$^{-1}$.

All data were reduced and calibrated using standard procedures within the GILDAS software \citep{Guilloteau2000}. After subtracting a first-order baseline, the data were re-gridded to a 30$^{\prime\prime}$ pixel grid and converted to main beam brightness temperature ($T_{\text{mb}}$), assuming a main beam efficiency $\eta_{\text{mb}}=0.46$ and a beam filling factor of $f_{\text{b}}\sim1$.
The typical rms noise levels are 0.45\,K for $^{12}$CO and 0.25\,K $^{13}$CO and C$^{18}$O.
These datasets constitute the largest $^{12}$CO, $^{13}$CO and C$^{18}$O maps to date that combine moderate angular resolution, fully sampled grids, and low noise levels.

We obtained a part of the MWISP CO molecular cloud data for the regions $95^\circ < l < 115^\circ$ and $0^\circ < b < 5^\circ$, which partially covers the Snake\,III stellar complex.
The aim of this paper is to exploit these data, in conjunction with stellar catalogs, to investigate the co-evolution of stars and molecular clouds.

\subsubsection{Distances of Molecular Clouds }\label{subsubsec:2.2.2}

\begin{figure}
    \centering
    \includegraphics[width=1\linewidth]{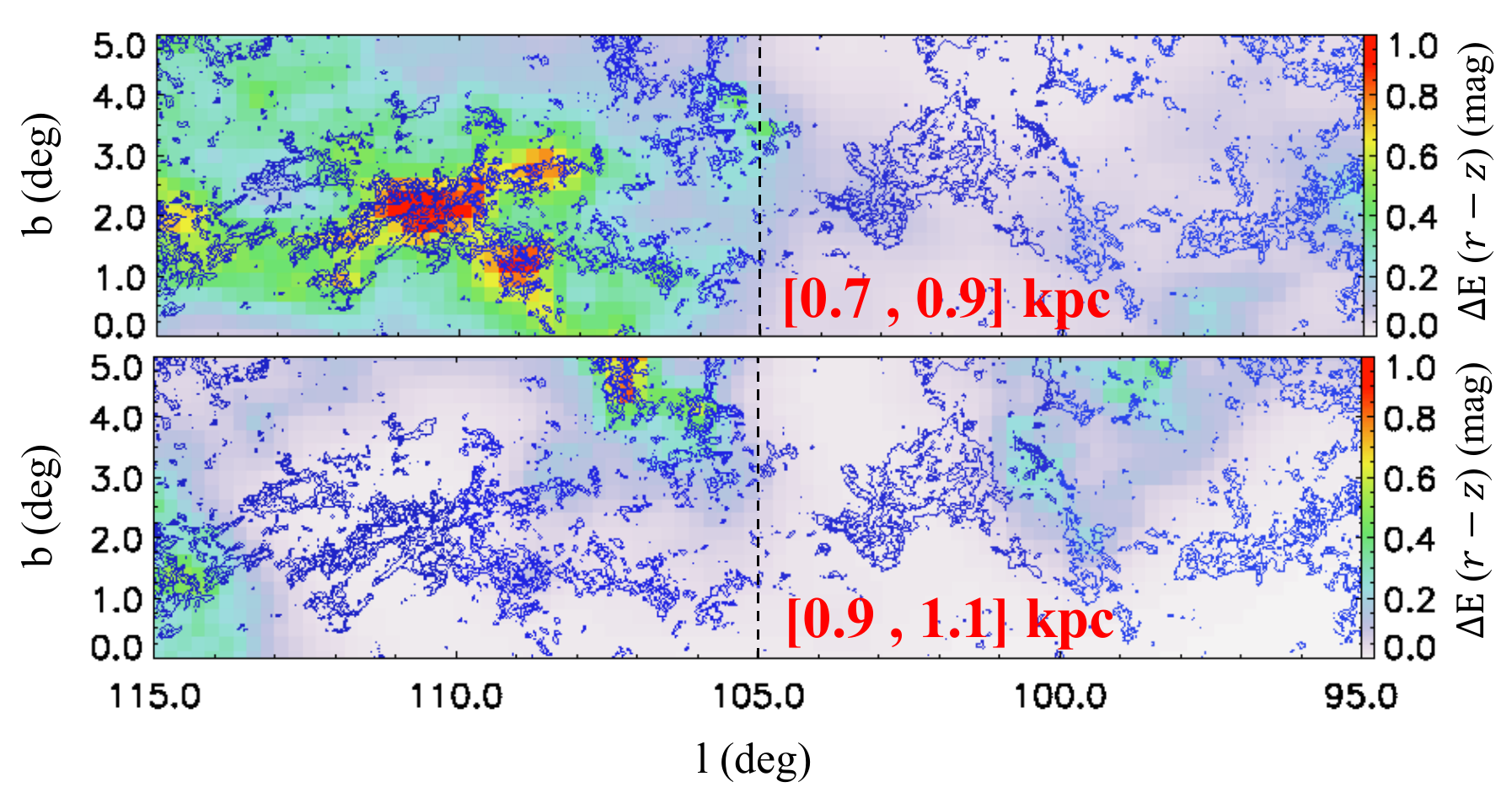}
    \caption{The background shows the Green 3D dust-extinction map integrated over 0.7--0.9\,kpc (upper) and 0.9--1.1\,kpc (lower). Overlaid are the MWISP $^{12}$CO integrated-intensity contours for integration range of [-20,5]\,km\,s$^{-1}$.}
    \label{fig.6}
\end{figure}

For molecular gas data, the resolution of cloud distances is of paramount importance.
We first integrated the 3-D dust-extinction map released by \citet{Green2019} along the line of sight across the Snake\,III stellar region observed by MWISP ($95^\circ < l < 115^\circ$ , $0^\circ < b < 5^\circ$), adopting distance bins of 700--900\,pc and 900--1100\,pc, which also represents the approximate distance range for Snake\,III region.
Overlaying the integrated intensity contours of  MWISP $^{12}$CO reveals that (Figure \ref{fig.6}), most of the molecular clouds on the left-hand side ($l > 105^\circ$) are found to coincide with the 700--900\,pc distance layer, matching the distance interval of the stars in that area. 
Between 900 and 1100\,pc only few clouds fragment at $l \approx$ $114.5^\circ$, $107^\circ$ and $99^\circ$ are present, which is reasonable: the clouds at $l \approx$ $114.5^\circ$, $107^\circ$ can be regarded extensions of the 700--900\,pc layer, while the stars near $l \approx$ $99^\circ$ are positionally coincident with scant clouds at that distance, so the presence of a minor cloud component is not unexpected. 
Therefore, we conclude that molecular gas associated with the Snake\,III stellar region should exhibit pronounced crowding at Galactic longitudes $l > 105^\circ$ and conspicuous sparsity at $l < 105^\circ$.

Subsequently, using the second Galactic quadrant molecular cloud distance catalog released by \citet[][{hereafter, Y21}]{Yan2021}, we obtain high-precision distance anchors across portions of the Snake\,III stellar region ($105^\circ < l < 115^\circ$ , $0^\circ < b < 5^\circ$), and identified the molecular clouds in this region as the Cep\,OB3 complex.
Based on The background-eliminated extinction-parallax(BEEP) Method, the catalog furnishes, at the 5\% precision level, reliable distances spanning 211--2631\,pc for 76 medium-scale molecular clouds for the first time.
Most notably, Y21 uncover a coherent giant molecular cloud complex, G125.1+02.6, stretching roughly 700\,pc in longitude and located at 1\,kpc, with a mass of 1.5$\times$10$^7$\,M$_\odot$, thus providing a new contextual backdrop to interpret the bulk kinematics of molecular gas in the Snake\,III region.

\begin{figure*}
    \centering
    \includegraphics[width=1\linewidth]{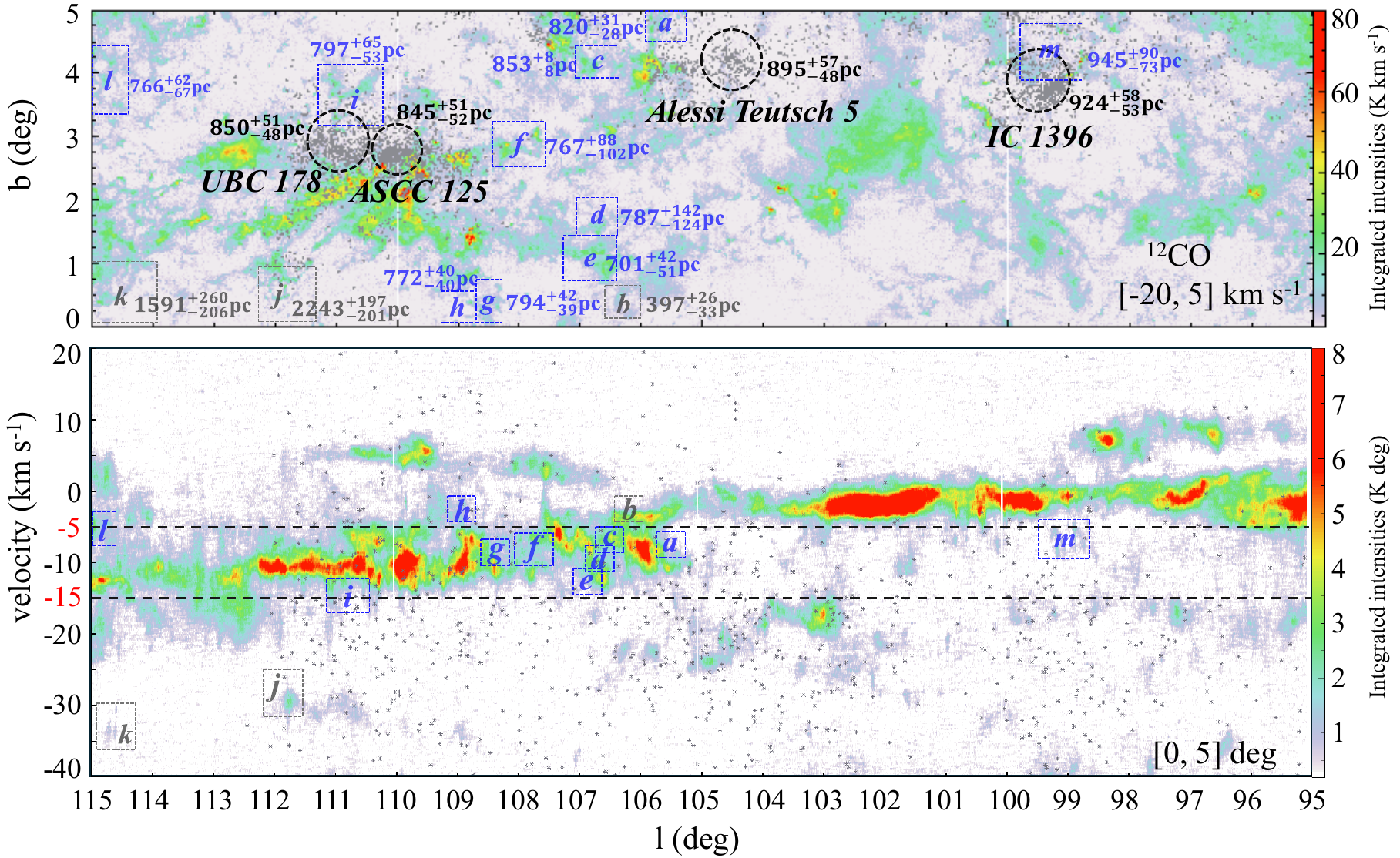}
    \caption{Galactic longitude-latitude map of $^{12}$CO integrated intensity over velocity (upper panel) and longitude–velocity map of $^{12}$CO integrated intensity over Galactic latitude (lower panel) overlaid with Snake\,III member stars (gray dots). Some molecular cloud distance and velocity released by the BEEP method marked as $a$, $b$, $c$, $d$, $e$, $f$, $g$, $h$, $i$, $j$, $k$, and $l$; and the IC\,1396 H\,II region marked as $m$. The four open clusters embedded within the molecular environment are circled in black in the upper panel; circle sizes scale with cluster radius, and their median distances with uncertainties are annotated.}
    \label{fig.7}
\end{figure*}

The portion of the Snake\,III star-formation region ($105^\circ < l < 115^\circ$ , $0^\circ < b < 5^\circ$) also encompasses half of the MWISP observational data that we have obtained.
By comparing the MWISP observational data with the molecular cloud distance catalog from {Y21}, we find some clouds in this region.
These clouds are G105.2+05.0, G106.1+00.5, G106.5+04.0, G106.5+01.6, G106.6+01.0, G107.7+02.9, G108.4+00.3, G109.0-00.1, G110.9+03.5, G111.7+00.0, G114.5-00.1 and G117.0+03.7.
In the integrated intensity map of $^{12}$CO shown in Figure \ref{fig.7}, they are marked as $a$, $b$, $c$, $d$, $e$, $f$, $g$, $h$, $i$, $j$, $k$, and $l$.
Except for clouds $b$, $j$, and $k$ (shown in gray in Figure \ref{fig.7}), all other clouds are closely correlated with the distances of Snake\,III member stars, falling within 700--900~pc.
In the right-hand region ($l < 105^\circ$), \citet{Pelayo2023} identified the IC\,1396 H\,II region \citep[also known as S131;][]{Salpeter1955} at a distance of 945$^{+90}_{-73}$~pc, which is part of the large, star-forming Cepheus bubble \citep{Patel1998}, and we marked this region as $m$ in Figure \ref{fig.7}.
It is also highly consistent with the stellar population associated with Snake\,III.
Beyond this, no additional molecular clouds that satisfy the criteria were identified in the right-hand region. 
In Figure \ref{fig.7}, we mark in black the four clusters identified throughout the region, along with their distances.

\subsubsection{Radial Velocities of Molecular Clouds}\label{subsubsec:2.2.3}

After confirming the molecular cloud distances, we still need to verify their radial velocities to strip clouds that are genuinely associated with the Snake\,III stellar region.
In fact, Y21 also measured the velocities of the medium-scale molecular clouds, allowing us to establish the velocities of the clouds at the boundary of the Cep\,OB3 complex.
According to the molecular cloud distance catalog, the velocities of the clouds identified in Section \ref{subsubsec:2.2.2} are -8.4, -2.9, -7.0, -10.8, -11.7, -9.9, -9.8, -2.8, -16.6, -29.9, -36.2 and -5.7\,km\,s$^{-1}$, and they are marked with the same letters and colors in the lower panel of Figure \ref{fig.7}.

It is readily apparent that, excluding clouds $b$, $j$, and $k$, the velocities of the remaining molecular clouds are concentrated mainly between -15 and -5\,km\,s$^{-1}$, with the IC\,1396 H\,II region also falling within this range.
Therefore, adopting an integration range of -15 to -5\,km\,s$^{-1}$ for the integrated intensity map of $^{12}$CO is well justified: it directly reveals the main structure of the Cep\,OB3 complex and is fully consistent with the median {$V_r$} of Snake\,III.
By jointly analyzing distance and velocity information, we can cleanly strip the Cep\,OB3 complex and IC\,1396 H\,II region from the MWISP molecular cloud data, providing a solid foundation for subsequent analyzes.

We also calculated the central velocity ($V_{\text{cen}}$) and velocity dispersion ($\sigma_v$) of the molecular clouds, adopting the interval range of [-15, -5]\,km\,s$^{-1}$.
To separate target clouds from background emission, we excluded any components whose $^{12}$CO integrated intensity (calculated in Section \ref{subsubsec:2.2.2}) falls below 2.5\,K\,km\,s$^{-1}$, thus removing noise contaminants from the centroid-velocity and velocity dispersion derived.
This filtering facilitates our subsequent investigation of star-cloud interactions.

\subsubsection{Excitation Temperature, Column Density and Mass of Molecular Clouds}\label{subsubsec:2.2.4}

For molecular cloud data, excitation temperature, column density, and mass are equally critical quantities of interest.

Assuming that the molecular clouds exist in LTE, we could obtain the excitation temperature ($ T_{\text{ex}}$) with the $^{12}$CO (J = 1--0) emission line, which is generally known to be optically thick, with $T^*_{\text{mb}}(^{12}\text{CO})$ integrated in the velocity range of [-15, -5]\,km\,s$^{-1}$.
Peaks with $ T_{\text{ex}}$ $>$ 15~K are evident in the centers of all cloud regions, indicating the presence of intense heating sources in the immediate vicinity of the molecular gas.

Because the data from $^{13}$CO and C$^{18}$O are not the focus of this article, the column density of H$_2$ is derived only from the integrated intensity of $^{12}$CO with the $X$-factor~=~1.8~$\times$~10$^{20}$\,cm$^{-2}$\,(K\,km\,s$^{-1}$)$^{-1}$ \citep{Dame2001}.

\begin{figure*}
    \centering
    \includegraphics[width=1\linewidth]{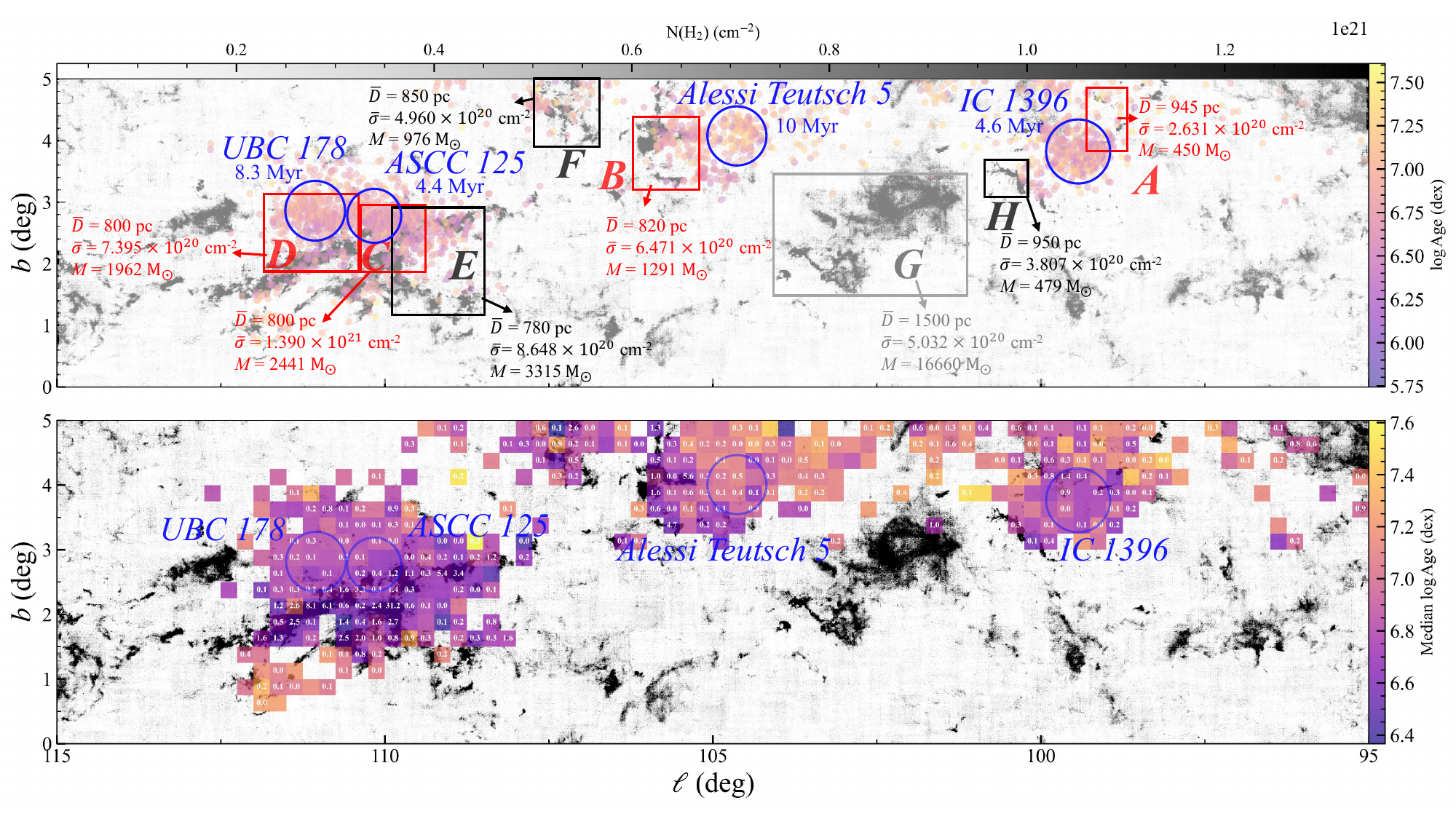}
    \caption{Upper panel: H$_2$ column density background map with stellar age $l-b$ distribution plot; the color bar shows the distribution of single-star age estimates fitted using Sagitta ($pms$ $>$ 0.1). Lower panel: H$_2$ column density background map with local stellar age $l-b$ distribution plot; the color bar shows the distribution of median local stellar age estimates fitted using Sagitta ($pms$ $>$ 0.1), and the rectangular bins have a side length of 0.25$^\circ$, with the local average column density marked at the center of each bin. The boxes mark several major regions and the average column density and total mass of the cloud within each region estimated by us. The red boxes mark molecular cloud regions (A, B, C, D) that host nearby open clusters, the black boxes indicate clouds (E, F, H) without adjacent open clusters, and the gray box denotes molecular cloud (G) at a different distance that should be distinguished from the others. We also indicate the positions of the four open clusters (blue circles) with their age estimates.}
    \label{fig.8}
\end{figure*}

Although there are significant cloud-to-cloud variations of the $X$-factor in the Galaxy \citep{Dame2001,Bolatto2013,Heyer2015}, we assume that this value is constant within one cloud.
The column density map of H$_2$ is shown in Figure \ref{fig.8}. 
We delineated several major regions and estimated the average column density and total mass of the cloud within each region, and the boundaries of each box in ($l$, $b$) are as follows:\\
Region\,A ($98.6^\circ<l<99.3^\circ$, $3.8^\circ<b<4.9^\circ$);\\ 
Region\,B ($105.2^\circ<l<106.2^\circ$, $3.2^\circ<b<4.4^\circ$); \\
Region\,C ($109.5^\circ<l<110.5^\circ$, $1.8^\circ<b<2.9^\circ$); \\
Region\,D ($110.5^\circ<l<111.9^\circ$, $1.8^\circ<b<3.0^\circ$); \\
Region\,E ($108.4^\circ<l<110.0^\circ$, $1.2^\circ<b<2.8^\circ$); \\
Region\,F ($106.7^\circ<l<107.7^\circ$, $3.9^\circ<b<5.0^\circ$); \\
Region\,G ($101.1^\circ<l<104.1^\circ$, $1.5^\circ<b<3.5^\circ$); \\
Region\,H ($100.1^\circ<l<100.9^\circ$, $3.0^\circ<b<3.7^\circ$).

\subsection{LAMOST MRS-N Observation}\label{subsec:2.3}

To gain a clearer picture of the co-evolution of stars and molecular clouds, we obtained a subset of observations from the LAMOST Medium-Resolution Spectroscopic Survey of Galactic Nebulae \citep[LAMOST MRS-N;][]{Wu2021}.
The survey employs the LAMOST telescope at the National Astronomical Observatories, conducting integral field spectroscopy at a resolution of R $\approx$ 7500 in two wavelength ranges: 4950--5350\,\AA (blue) and 6300--6800\,\AA (red).
Each exposure lasts 900\,s and every 5$^\circ$-diameter field (20\,deg$^2$) is observed 16 times, resulting in a total integration time of 4\,h.
The survey covers the northern Galactic plane strip that spans $40^\circ < l < 215^\circ$ in longitude and$ -5^\circ < b < 5^\circ$ in latitude ($\approx$ 1700~deg$^2$), with additional intensive observations of the four large nebulae Rosette, Westerhout\,5, Cygnus Loop, and Simeis\,147.
After initial dark-current subtraction, flat-field correction, spectral extraction, and first-order wavelength calibration by the LAMOST 2-D pipeline, the data are processed through an independent pipeline developed by the MRS-N team for cosmic-ray removal, sky-emission-line fitting and subtraction, second-order wavelength calibration, and emission-line parameter measurement.
Following recalibration with seven sky emission lines by \citet{Ren2021}, the radial velocity precision is better than 1\,km\,s$^{-1}$.

In a portion of the Snake\,III stellar region ($105^\circ < l < 115^\circ$ , $0^\circ < b < 5^\circ$), we obtained two epochs of LAMOST MRS-N observations.
After imposing a signal-to-noise cut (S/N of H$_\alpha$ $>$ 10) and removing stellar contributions, there are 6214 emission line detections and computed its relative integrated intensity after sky-line calibration at 6553~\AA.

\section{Result}\label{sec:3}

Building on the stellar and molecular-cloud quantities derived in the previous section (integrated intensities, excitation temperature, central velocity and H$_\alpha$ emission), we present $l-b$ comparison maps for Region A (without H$_\alpha$ emission) in Figure \ref{fig.10} and for Regions B, C, and D in Figure \ref{fig.11} (with H$_\alpha$ emission).
These maps reveal several morphological coincidences between stars and molecular clouds.

\subsection{Correlation between Stellar Age and Gas Density}\label{subsec:3.1}

The MWISP molecular-cloud data ($95^\circ< l < 115^\circ$, $0^\circ< b<5^\circ$) encase the four clusters IC\,1396, Alessi\,Teutsch\,5, ASCC\,125, and UBC\,178, each of which is still embedded in a residual molecular envelope, as well as extensive structures containing field stars.
In Sections \ref{subsubsec:2.1.4} and \ref{subsubsec:2.2.3}, we calculated the ages of the four clusters (using Sagitta and ASteCA) and the field stars in the extended structures (using Sagitta), along with the column densities and masses of the nearby major molecular clouds.
By combining these results, we can quantitatively investigate the correlation between the ages of clusters and field stars and the gas densities.

\subsubsection{Cluster Ages and Gas Density}\label{subsubsec:3.1.1}

Figure \ref{fig.8} presents the column density and mass estimates for the four molecular-cloud regions A, B, C, and D.
We find a global trend of increasing the gas density towards higher Galactic longitude (Regions C and D share the same molecular complex, with C denser than D).
Thus, the density gradient A$\rightarrow$B$\rightarrow$C+D is preserved.

Examination of the nearest clusters to these four regions reveals that: ASCC\,125, near the highest-density Region C, is extremely young ($\sim$4.4\,Myr), whereas UBC\,178, embedded in the same complex but near the slightly lower-density Region D, is relatively old ($\sim$8.3\,Myr); 
Region B also exhibits relatively high density, and the nearby cluster Alessi Teutsch 5 is the oldest ($\sim$10\,Myr); 
while IC\,1396, near the lowest-density Region A, is relatively young ($\sim$4.6\,Myr).
Therefore, the ages of these four star clusters can be roughly divided into two groups: old (UBC\,178 and Alessi\,Teutsch\,5) and young (ASCC\,125 and IC\,1396). 
Among them, the two older clusters are almost detached from the gaseous environment and located in gas-rarefied regions, while the two younger clusters still partially overlap with some gas.
Meanwhile, since the ages of these clusters are all less than or equal to 10\,Myr, their dynamical evolution is insufficient to cause the clusters to completely detach from the gaseous environment.
We therefore regard the overall trend that clusters formed earlier in regions near where the present-day gas density is higher as robust, while the exception presented by ASCC\,125 likely reflects a second-generation stellar population triggered by feedback from neighboring clusters or more.
A physical interpretation of this phenomenon will be presented in Section \ref{subsec:4.1}.

\subsubsection{Filed Star Ages and Gas Density}\label{subsubsec:3.1.2}

\begin{figure}
    \centering
    \includegraphics[width=1\linewidth]{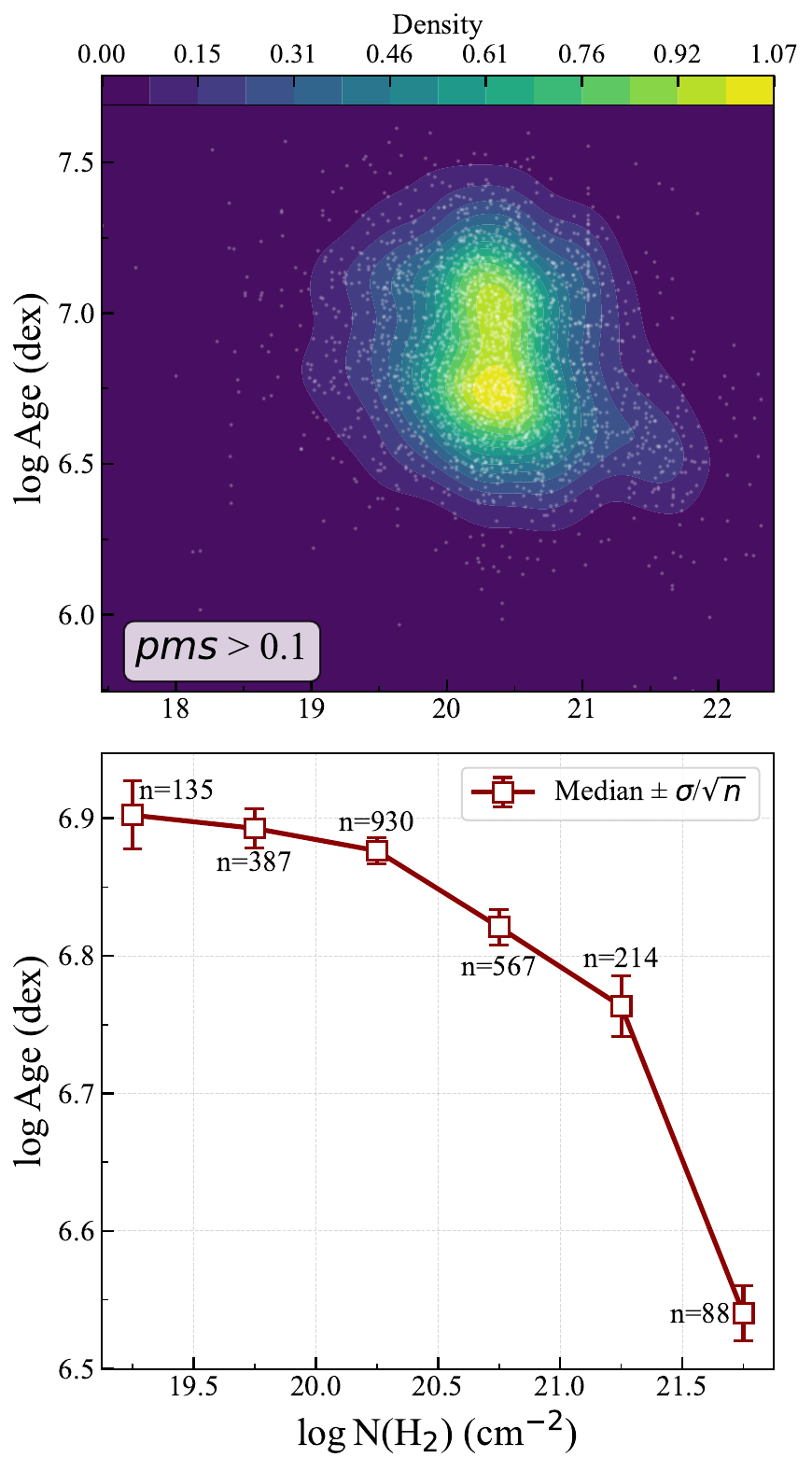}
    \caption{Contour plot (upper panel) and box plot distribution (lower panel) of Sagitta estimated stellar ages ($pms$ $>$ 0.1) versus local H$_2$ column density values at their positions; both age and H$_2$ column density are taken in logarithmic scale.}
    \label{fig.9}
\end{figure}

During the process of star formation and evolution, gas is continuously consumed, which is a dynamic process.
Based on the age estimates of thousands of stars fitted by Sagitta, we will be able to capture this dynamic process, thereby quantitatively studying the correlation between stellar ages and local gas densities in the present-day gas environment.

To this end, we have compiled the ages distribution for all stars with $pms$ $>$ 0.1 fitted by Sagitta in the Galactic coordinate $l-b$ plane, with the H$_2$ column density map as background, as shown in Figure \ref{fig.8}.
We can intuitively see that some young field stars are forming in the high-density residual gas.
To demonstrate this trend more clearly, we extracted the H$_2$ column density values at the position of each star and combined them with the stellar age estimates to plot contour maps and age statistics binned by column density (Figure \ref{fig.9}).

Thus, we find that under the present-day gas density gradient, more young stars tend to be born in high-density gas environments and have not yet aggregated to form clusters in large numbers; 
whereas older stars exist in medium-low density gas environments and have mostly already merged into the four identified clusters.
This reveals the statistical result that as the gas density at the stellar location increases, the stellar age structure becomes progressively younger.

\subsection{The Link between Stars and Molecular Clouds}\label{subsec:3.2}

\begin{figure*}
    \centering
    \includegraphics[width=1\linewidth]{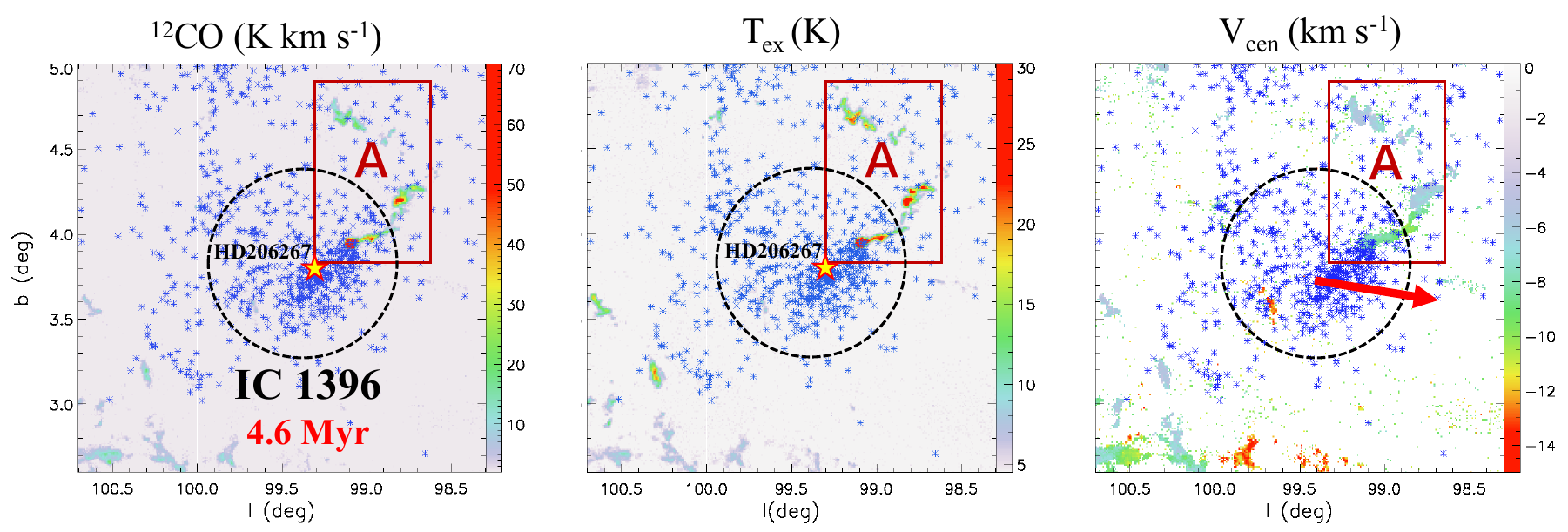}
    \caption{The $^{12}$CO integrated intensity, excitation temperature ($T_{\text{ex}}$) and centroid velocity ($V_{\text{cen}}$) of clouds in Region\,A and the scatter plot of IC\,1396 in Galactic coordinates ($l, b$). All molecular cloud-related maps use a velocity integration range of [$-15,-5$]\,km\,s$^{-1}$. The dark-red solid box marks the Region\,A molecular clouds. The black solid circle marks IC\,1396 (circle size scaled to the cluster radius). The red arrow indicates the median tangential velocity of IC\,1396. The yellow star symbol marks the multiple system (O6.5$+$B0$+$O8) HD\,206267.}
    \label{fig.10}
\end{figure*}

\begin{figure*}
    \centering
    \includegraphics[width=1\linewidth]{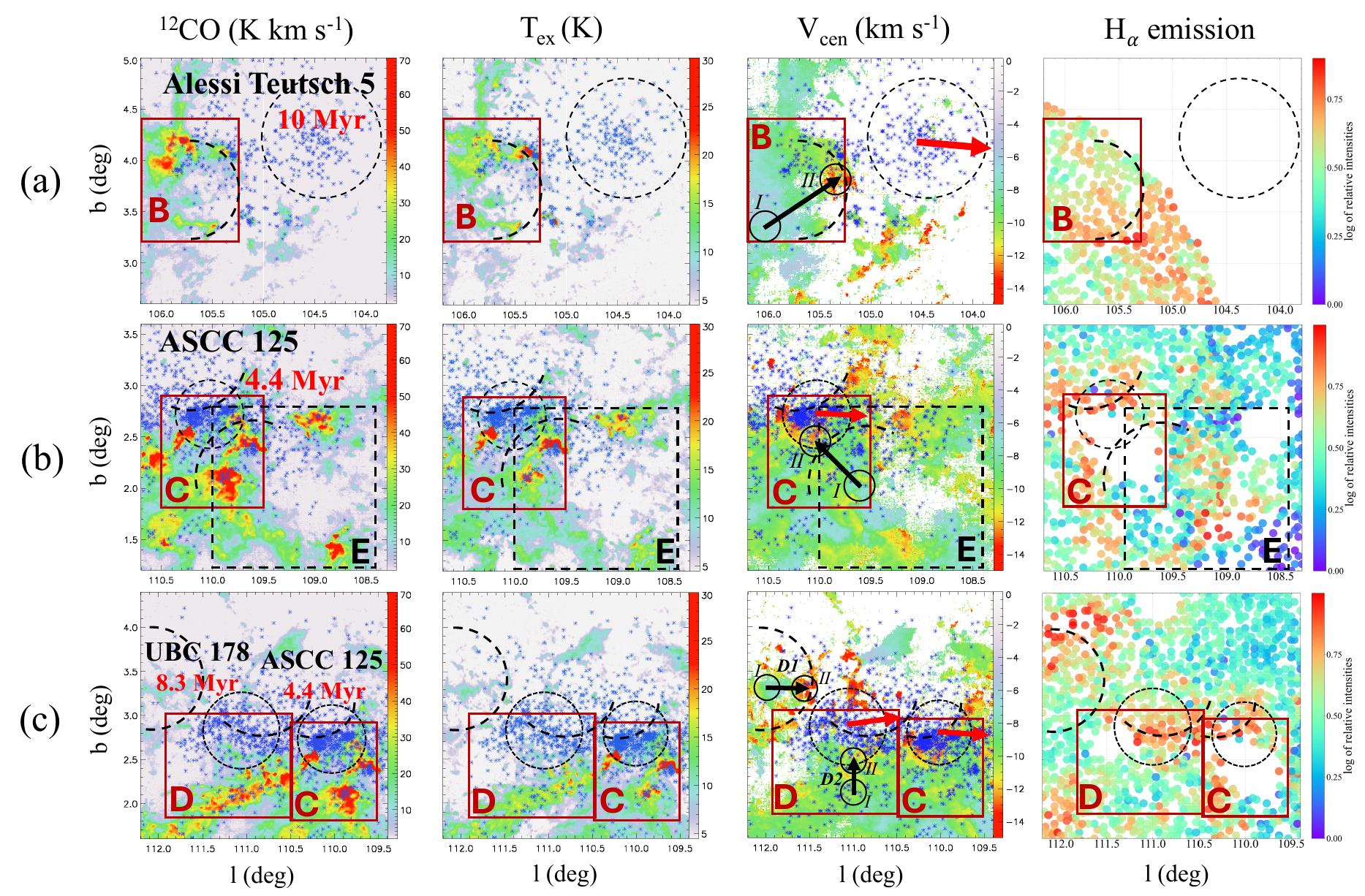}
    \caption{The $^{12}$CO integrated intensity, excitation temperature ($T_{\text{ex}}$) and centroid velocity ($V_{\text{cen}}$) of clouds in Regions\,B, C, and D, the scatter plot of stars and LAMOST MRA-N H$_\alpha$ emission in Galactic coordinates ($l,b$) among sub-panels (a), (b), and (c).
    All molecular cloud-related maps use a velocity integration range of [$-15,-5$]\,km\,s$^{-1}$.
    The dark-red solid boxes mark the Regions\,B, C, and D molecular clouds.
    The black solid box marks the Region\,E molecular clouds.
    The black dashed circles mark the open clusters (circle size scaled to the cluster radius).
    The red arrows indicates the median tangential velocity of corresponding clusters.
    The black arrows in (a) indicate the possible direction in which Alessi\,Teutsch\,5 detached from the molecular clouds in Region\,B, as well as the selected unperturbed area (I) and perturbed area (II) along this direction; the black arrows in (b) indicate the direction in which some as-yet-unidentified physical agent in Region\,E compresses the gas surrounding ASCC\,125 near Region\,C, with the unperturbed area (I) and perturbed area (II) selected; the black arrows in (c) indicate the direction in which UBC\,178 detached from its natal gas (D1) and the direction compressing the gas in Region\,D (D2), with unperturbed area (I) and perturbed area (II) selected respectively. The black dashed arcs depict the compressed or perturbed shells of the molecular clouds by stellar feedback.}
    \label{fig.11}
\end{figure*}

To investigate star-gas interactions, we combine the stellar and molecular cloud datasets.
Following the identification of spatially and kinematically associated clouds (Sections~\ref{subsubsec:2.2.2} and \ref{subsubsec:2.2.3}) and the derivation of their physical properties (Section~\ref{subsubsec:2.2.4}), we present the merged results in Figures~\ref{fig.10} and \ref{fig.11}. 
We divide the molecular clouds containing clusters into four Regions (A--D) and overlay the corresponding stellar populations, including four open clusters: IC\,1396, Alessi\,Teutsch\,5, ASCC\,125, and UBC\,178.
This enables us to examine spatial, kinematic, and stellar-heating correlations between stars and gas.
Figure~\ref{fig.10} shows Region~A's stellar distribution overlaid on the $^{12}$CO integrated intensity, excitation temperature, and centroid-velocity maps (no MRS-N data available).
Figure~\ref{fig.11} presents the same layout for Regions B--D, with additional H$_\alpha$ emission panels (calibrated to the 6553~\AA\ sky line).

As shown in Figure \ref{fig.10}, Region\,A, the molecular cloud around the stars is tenuous, and the open cluster IC\,1396 is located here.
Numerous earlier studies conclude that this cluster was born within the IC\,1396 H\,II region \citep[also designated S131;][]{Sharpless1959}, a large star-forming Cepheus bubble ionized by the multiple system (O6.5$+$B0$+$O8) HD\,206267 \citep{Harvin2003,Saurin2012} and described by \citet{Patel1998}.
The $^{12}$CO integrated-intensity map reveals several bright-rimmed clouds (BRCs), consistent with earlier studies \citep{Patel1995,Barentsen2011}; these BRCs are designated IC\,1396A, IC\,1396B and IC\,1396N \citep[labeled “E” by][]{Pottasch1956,Patel1995}.
The excitation-temperature map shows that the BRCs are heated to $ T_{\text{ex}}$ $>$ 20~K, further demonstrating the heating of molecular clouds by stellar winds.
Centroid-velocity images reveal uniformly layered, quasi-quiescent velocity fields throughout the clouds.

In Figure \ref{fig.11}(a), a portion of the molecular clouds (Region\,B) remain on the western side of the open cluster Alessi\,Teutsch\,5 .
The BRC adjacent to the cluster exhibits strong heating, with excitation temperatures $ T_{\text{ex}}$ $>$ 20~K, and its centroid velocity shows pronounced perturbations, which we attribute to the influence of Alessi\,Teutsch\,5 itself.
The H$_\alpha$ emission is faint and cannot be reliably separated from the background; moreover, no MRS-N data are available for the stellar region.

And in Figure \ref{fig.11}(b), the molecular cloud in Region\,C presents a bowl-shaped morphology, with the highly concentrated cluster ASCC\,125 embedded within the “bowl”.
Numerous BRCs are heated to nearly 30~K, and the cloud’s centroid velocity shows strong perturbations and clear signs of expansion.
Comparing the H$_\alpha$ emission, we find that Region\,C appears to be part of a larger ring-like H$_\alpha$ structure on its lower-right side; we mark this structure as Region\,E, which overlaps with Region\,C.
The $^{12}$CO integrated-intensity map also reveals a molecular cloud bubble that is about 1$^\circ$ in radius, seemingly the source of the surrounding ring-like H$_\alpha$ emission.
The origin of this bubble is unknown, but it may have influenced star formation in the vicinity.

As shown in Figure \ref{fig.11}(c), the molecular cloud in Region\,D exhibits a disk-shaped morphology, with the dense cluster UBC\,178 positioned above the "disk".
The BRCs within the disk are heated to $ T_{\text{ex}}$ $>$ 20\,K, and the clouds' centroid velocity is perturbed and expanding, enveloping the cluster.
A large-scale perturbation in centroid velocity is observed in the molecular cloud to the left of the cluster, likely attributable to the early activity of UBC\,178.
The H$_\alpha$ emission is consistent with the cluster acting as the ionizing source.

\subsection{Velocity Dispersion of Local Molecular Cloud}\label{subsec:3.3}

\begin{figure*}
    \centering
    \includegraphics[width=1\linewidth]{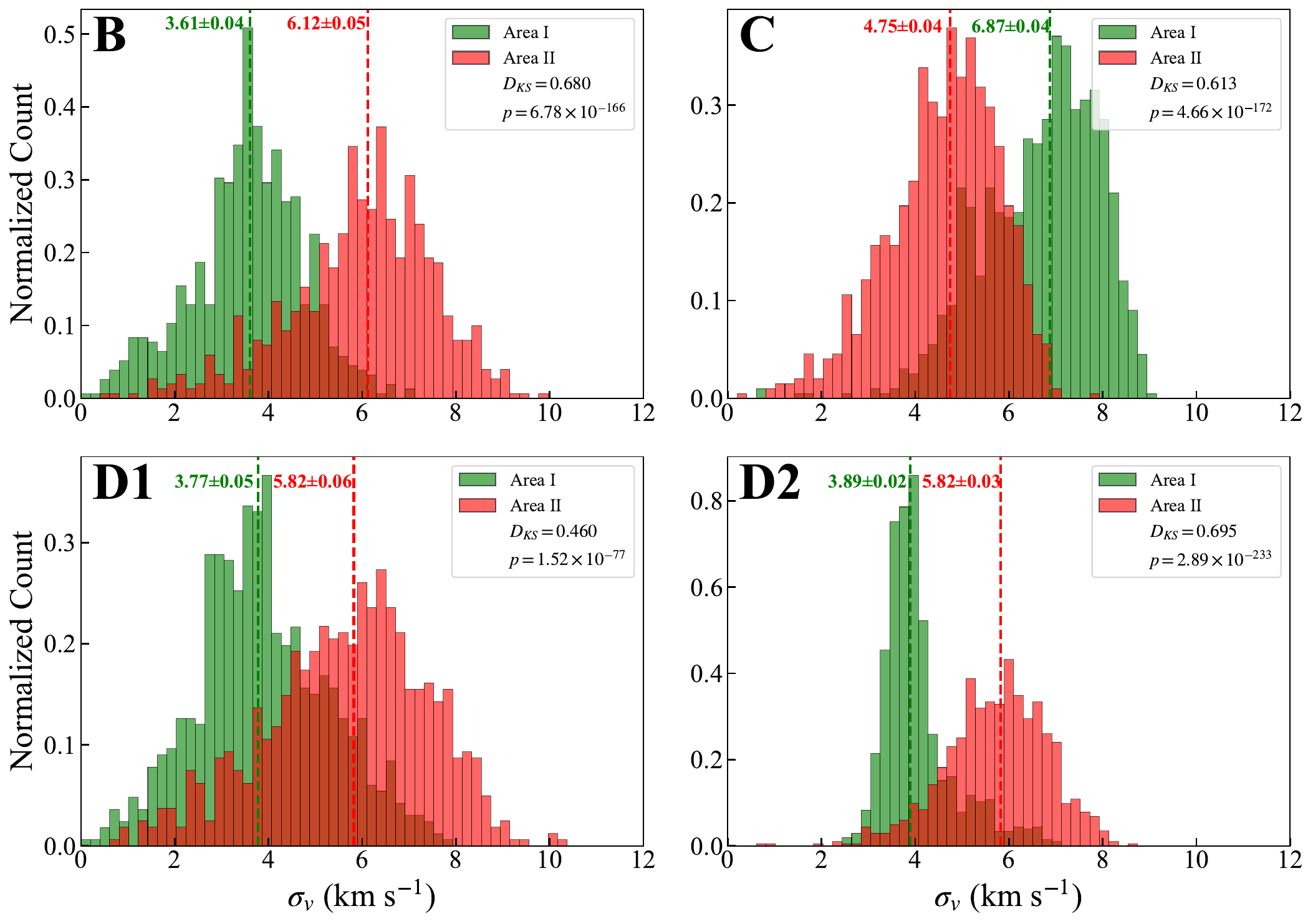}
    \caption{Histogram comparing the velocity dispersion ($\sigma_v$) distributions selected in Regions B, C, D. Area\,I (green, unperturbed) and Area\,II (red, perturbed) are plotted in different colors, with their median $\sigma_v$ values and uncertainties calculated respectively.
    {The $D_{KS}$ and $p$ values shown in the label represent the K-S tests performed for each comparison region, all demonstrating the significance of the difference in $\sigma_v$ distributions between Area\,I and Area\,II ($p\ll0.001$).}
    Note that the y-axes of all plots use normalized counts with different ranges.}
    \label{fig.12}
\end{figure*}

In order to quantify the velocity perturbations of the molecular cloud caused by suspected stellar feedback near Regions B, C, and D observed in the centroid-velocity map shown in Figure \ref{fig.11}, we selected two areas along each of the black arrow directions in the figure to calculate their velocity dispersion ($\sigma_v$).
The resulting histogram of the $\sigma_v$ distributions is shown in Figure \ref{fig.12}.
{By quantitatively comparing $\sigma_v$ between cluster-proximate perturbed area and the cluster-distal unperturbed area (which in Region\,C appears as the perturbed area affected by the E-bubble) using the Kolmogorov–Smirnov (K-S) test, and comparing $\sigma_v$ with the sound speed of the CO molecular cloud, we will be able to quantify the degree to which stellar feedback perturbs the molecular cloud.}
We adopt the typical sound speed of $c_s = \sqrt{k_B T / \mu m_H} \approx 0.25\sim0.29$\,km\,s$^{-1}$, corresponding to $^{12}$CO molecular cloud at $T$=15 $\sim$ 20\,K, to estimate the Mach number ($\mathcal{M} = \sigma_v / c_s$) for different regions \citep{McKee2007}.

For Region\,B, along the direction in which Alessi\,Teutsch\,5 was drifting away, the centroid velocity of the molecular clouds is strongly perturbed, and the $\sigma_v$ of the perturbed Area II (6.02$\pm$0.05\,km\,s$^{-1}$) and the unperturbed Area I (3.61$\pm$0.04\,km\,s$^{-1}$) shown in Figure \ref{fig.12} are clearly separated.
The Mach number $\mathcal{M}$ reaches 20.76 $\sim$ 24.08 in the perturbed area and 12.45 $\sim$ 14.44 in the area region, indicating that the gas is in a highly supersonic turbulent state.
Meanwhile, some young field stars appear to be aggregating to form cluster cores near the perturbed area.

For Region\,C, by measuring the $\sigma_v$ of the area perturbed by ASCC\,125 ({Area\,II}) and the area with the highest gas density below ({Area\,I}, with a local average density value reaching 3$\times$10$^{22}$\,cm$^{-2}$, as shown in Figure \ref{fig.8}), we found that the $\sigma_v$ in the {Area\,II} is $\sigma_v$ = 4.75$\pm$0.04\,km\,s$^{-1}$, with $\mathcal{M}$ = 16.38 $\sim$ 19.00, which is slightly lower than that of the D2 perturbed area but slightly higher than the D2 unperturbed area.
Meanwhile, the $\sigma_v$ in the {Area\,I} is $\sigma_v$ = 6.87$\pm$0.04\,km\,s$^{-1}$, with $\mathcal{M}$ = 23.69 $\sim$ 27.48, reaching the highest value and indicating some form of superb feedback activity to the right of Region\,C {(highly likely to be E-bubble)}.

For Region\,D, we measured the velocity dispersion \(\sigma_v\) in the perturbed (Area~II) and unperturbed (Area~I) areas along two directions, D1 and D2, relative to the cluster UBC\,178 (Figure~\ref{fig.12}). 
For D1, Area~II yields $\sigma_v$ = 5.82$\pm$0.06\,km\,s$^{-1}$ (\(\mathcal{M} = 20.07\)$\sim$\(23.28\)), while Area~I gives $\sigma_v$ = 3.77$\pm$0.05\,km\,s$^{-1}$ (\(\mathcal{M} = 13.00\)$\sim$\(15.08\)).
For D2, Area~II shows $\sigma_v$ = 5.82$\pm$0.03\,km\,s$^{-1}$ (\(\mathcal{M} = 20.07\)$\sim$\(23.28\)), and Area~I gives $\sigma_v$ = 3.89$\pm$0.02\,km\,s$^{-1}$ (\(\mathcal{M} = 13.41\)$\sim$\(15.56\)).
Notably, despite D1 and D2 representing different directions relative to the cluster, the velocity dispersions in the perturbed regions are remarkably consistent, indicating that the molecular gas in these areas is in a highly supersonic turbulent state.

\section{Discussion}\label{sec:4}

Snake\,III exhibits strong evidence of primordial physical homogeneity. Its member stars are distributed across an extensive spatial volume and display a median radial velocity of \(-20.57 \pm 0.51\,\text{km}\,\text{s}^{-1}\), along with tightly coherent tangential velocity distributions. This kinematic coherence suggests a common origin and evolutionary history, largely unaffected by external perturbations.

Moreover, the molecular clouds spatially associated with Snake\,III member stars show excellent agreement in both distance and velocity with the stellar population, confirming that these clouds are the birth sites of the stars. The interaction between stars and molecular clouds is particularly pronounced in high-density regions, reinforcing the interpretation of Snake\,III as a primordial structure formed within Galactic molecular clouds.

These findings offer valuable insights into the formation and evolution of complex stellar structures in the Milky Way and contribute to our understanding of the interplay between star formation and molecular clouds.

In the following, we delineate a coherent formation scenario for the global star-cloud complex and provide a dedicated examination of the origin of ASCC\,125.

\subsection{Overall Evolutionary Picture}\label{subsec:4.1}

Based on the observational trends presented in Section~\ref{sec:3}—namely (i) a monotonic increase in molecular cloud density along the sequence A→B→C+D, (ii) a systematic decrease in stellar age with increasing present-day gas density, and (iii) clear kinematic and thermal signatures of stellar feedback (e.g., enhanced line widths, elevated excitation temperatures, and localized bubbles)—we now synthesize a unified formation scenario for the Snake\,III complex. Before doing so, we state two necessary assumptions:
\begin{enumerate}
    \item As clusters evolve, they gradually detach from their parent gas through consumption and dispersal. UBC\,178 and Alessi\,Teutsch\,5 already reside in gas‑rarefied cavities, while ASCC\,125 and IC\,1396 still partially overlap with their natal material but show signs of separation. Hence, when correlating cluster age with gas density, we consider the nearest‑neighbour gas regions as remnants of the original molecular clouds rather than requiring exact spatial coincidence.
    \item The present‑day (residual) gas density reflects, at least qualitatively, the density gradient of the parent molecular clouds at the time of cluster formation.
\end{enumerate}

These assumptions imply that the A→B→C+D density gradient was already imprinted in the initial cloud complex. Theoretical work (\citealt{Krumholz2005, Hennebelle2008}) shows that higher densities lower the Jeans mass and promote efficient massive star formation. Accordingly, the initially densest parts (near B and D) formed stars earlier, producing the relatively old clusters Alessi\,Teutsch\,5 and UBC\,178, which have since largely cleared their surroundings. In the lower‑density A region, star formation was delayed and required external triggering, resulting in the younger IC\,1396. The present‑day dense gas (especially in C) continues to form young field stars, indicating that star formation is still active there.

Stellar feedback then profoundly reshapes the remaining gas. In Region\,A, the multiple system HD\,206267 (O6.5+B0+O8) has swept most of the gas into a ring, leaving a tenuous cavity (Figure~\ref{fig.10}). Three consequences follow: (i) the gas density is too low to form new massive cores; (ii) the radiation field is dilute and no potential well remains for assembling massive aggregates; (iii) the residual gas shows a quiescent velocity field. Star formation in IC\,1396 has therefore effectively ceased.

In Region\,B, the cluster Alessi\,Teutsch\,5 is drifting away from its parent cloud, carrying only a small fraction of gas with it (Figure~\ref{fig.11}a). The feedback has (i) increased the Jeans mass in the remaining gas, quenching further cluster formation, and (ii) strongly perturbed the gas along the cluster’s trajectory (velocity dispersion \(\sigma_v \approx 6\)\,km\,s\(^{-1}\), Mach number \(\mathcal{M}\sim20\)–24). This perturbation has locally compressed the gas, triggering the formation of young field stars that are now aggregating into new cluster cores.

Region\,C hosts the extremely young (4.4\,Myr) cluster ASCC\,125, which sits inside a prominent CO cavity (Figure~\ref{fig.11}b). Its intense radiation heats the surrounding gas (\(T_{\mathrm{ex}} > 20\)\,K) and perturbs the kinematics (\(\sigma_v \approx 4.75\)\,km\,s\(^{-1}\)). The even higher velocity dispersions measured in the adjacent D2 {Area\,II} and in the dense gas below C ({Area\,I}) indicate that ASCC\,125 formed from material compressed by feedback from both UBC\,178 and an additional agent ({possibly the E‑bubble}). Hence ASCC\,125 represents a second‑generation stellar system, born in a high‑density environment that was itself created by earlier feedback.

Region\,D contains a dense, disk‑shaped molecular cloud irradiated by UBC\,178 (Figure~\ref{fig.11}c). The low‑density gas adjacent to the cluster exhibits strong velocity perturbations, while the FUV flux and stellar winds have carved shallow furrows along the cloud surface. UBC\,178 is older than ASCC\,125 and its proper motion points toward the disturbed cloud on the western side, suggesting that it formed in that now‑dispersing gas and subsequently compressed the clouds in Regions\,C and D, triggering a new generation of stars.

In summary, a coherent evolutionary picture emerges: three first‑generation open clusters (UBC\,178, Alessi\,Teutsch\,5, and IC\,1396) formed in progressively less dense portions of the original giant molecular cloud. Their feedback—through winds, ionizing radiation, and possibly supernovae in E-bubble—sculpted the remaining gas, compressing some regions to trigger second‑generation star formation (exemplified by ASCC\,125 and numerous field stars) while evacuating others. This interplay between initial cloud structure and subsequent feedback explains the observed correlations between stellar age, gas density, and cloud kinematics across the Snake\,III complex.

\subsection{The Formation of ASCC\,125}\label{subsec:4.2}

\begin{figure}
    \centering
    \includegraphics[width=1\linewidth]{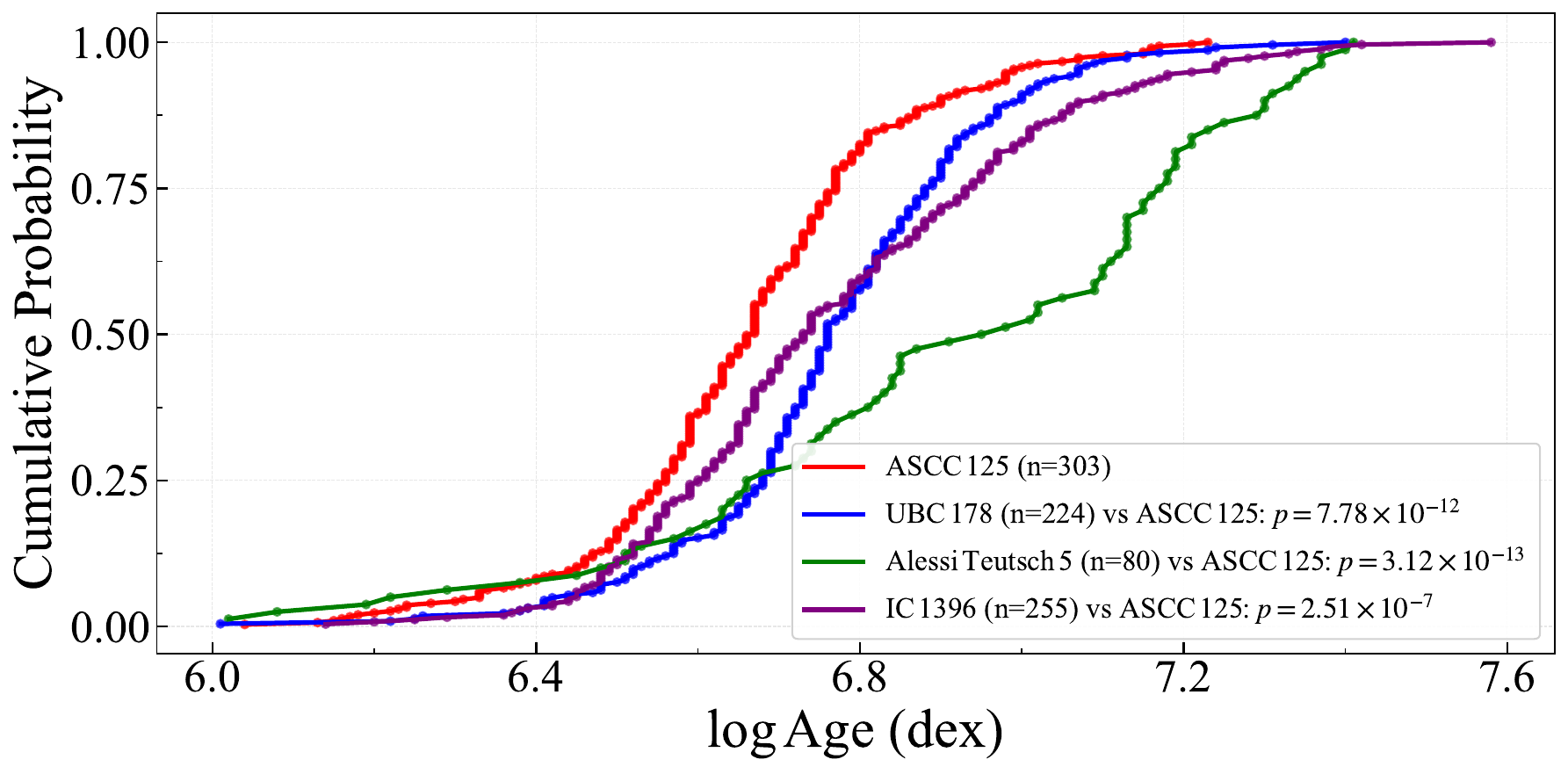}
    \caption{The cumulative distribution function (CDF) plots of Sagitta ages ($pms$ $>$ 0.1) for the four open clusters, with logarithmic age on the x-axis and cumulative probability (0–1) on the y-axis. The p-values from K-S tests between other clusters and ASCC\,125 are labeled in the figure to indicate its significance ($p\ll0.001$).}
    \label{fig.13}
\end{figure}

In light of the evolutionary picture outlined above, ASCC\,125 no longer appears as an anomaly; rather, its unusually young age reflects a history of exceptionally vigorous stellar feedback rather than peculiar initial conditions. To quantify this, we applied K-S tests to the single-star ages (Sagitta, \(pms > 0.1\)) of ASCC\,125 against those of UBC\,178, IC\,1396, and Alessi\,Teutsch\,5. As shown in Figure~\ref{fig.13}, the age distribution of ASCC\,125 is concentrated at the youngest end and differs significantly from all three comparison clusters (\(p \ll 0.001\)).

We note that the Sagitta age of UBC\,178 (\(\sim5.8\) Myr) is considerably younger than its ASteCA cluster age (\(\sim12\) Myr). This discrepancy likely arises from spatial contamination by young stars recently formed in or near ASCC\,125, which are located in close proximity to UBC\,178. Hence a more robust age estimate for UBC\,178 is the average of both methods (\(\sim8.3\) Myr). In contrast, Alessi\,Teutsch\,5 is more isolated and shows good agreement between its Sagitta (\(\sim9.3\) Myr) and ASteCA (\(\sim10.7\) Myr) ages. Due to this environmental difference, the age difference between UBC\,178 and Alessi\,Teutsch\,5 is not actually significant, and both are classified as old clusters; therefore, the phenomenon that they formed in similar initial gas density environments is reliable. Although the absolute ages of UBC\,178 and Alessi\,Teutsch\,5 are similar, it remains clear that UBC\,178 is older than ASCC\,125, regardless of the adopted age scale. 

This quantitative analysis reinforces the interpretation of ASCC\,125 as a second‑generation cluster, born from gas that was compressed and triggered by the feedback of its more massive neighbour, UBC\,178, and possibly by the supernova that created the E‑bubble.

A plausible formation scenario for ASCC\,125 is as follows. Stellar winds from the older cluster UBC\,178 compressed the surrounding gas, locally enhancing the density and initiating star formation in the region that would become ASCC\,125. Concurrently, a burst of massive-star formation within the E-bubble generated intense feedback that rapidly evacuated the surrounding gas, carving a cavity in the parent cloud. This process may have culminated in a supernova explosion, whose expanding shell swept up gas and further compressed it into the bubble wall.

When this expanding shell encountered the ASCC\,125 region, it collided with the wind-driven front produced by UBC\,178. The resulting compression triggered a vigorous new episode of star formation, giving birth to the cluster ASCC\,125. Feedback from both sides drove gas to converge around the nascent cluster, enabling it to assemble into the dense stellar aggregate observed today, while field stars continue to form within the remaining gas reservoir.

\begin{figure}
    \centering
    \includegraphics[width=1\linewidth]{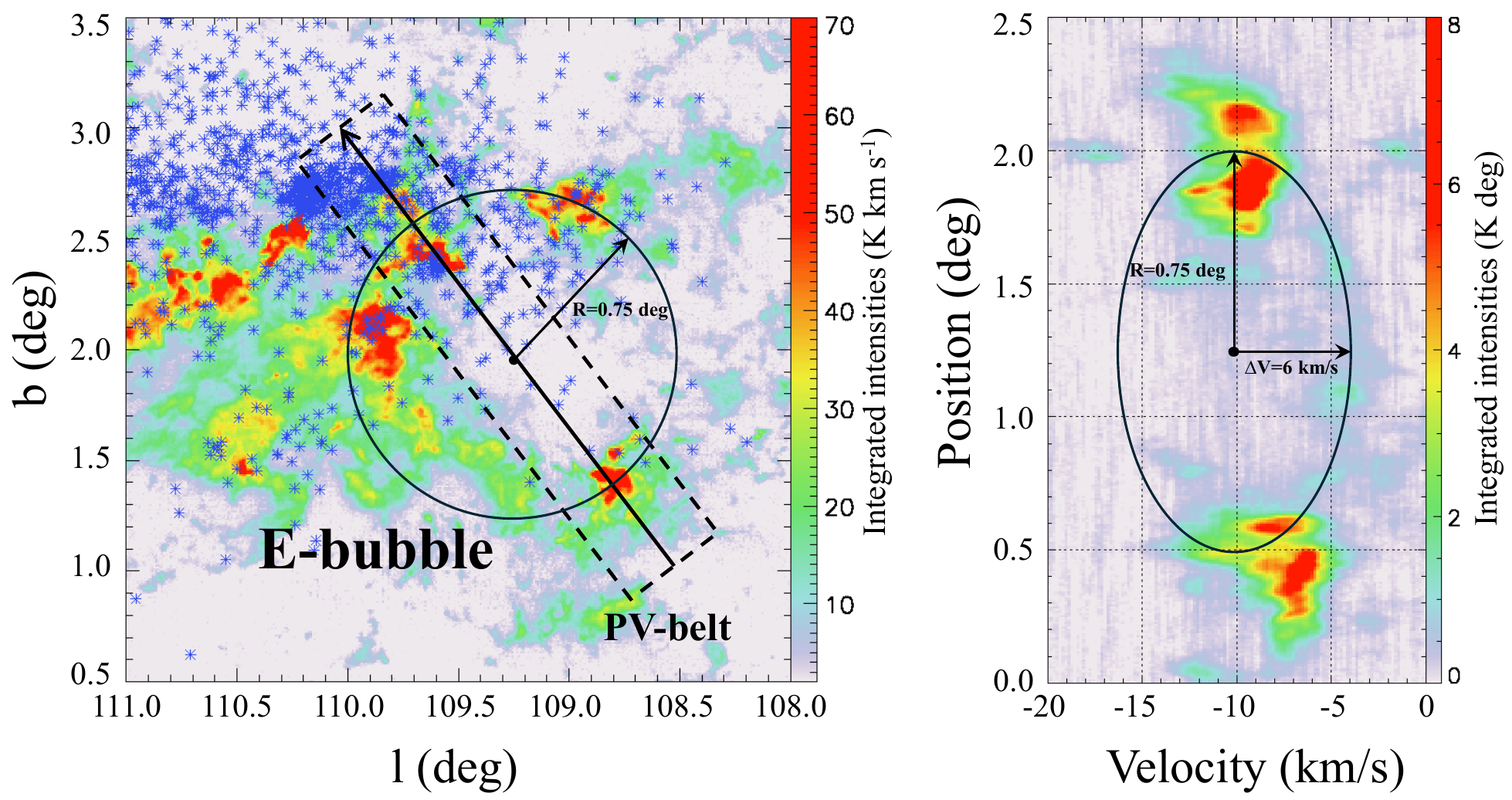}
    \caption{{The $^{12}$CO integrated intensity map of the bubble in Region\,E (left panel) and the position-velocity (P-V) diagram of the selected PV-belt (right panel).} In the left panel, the rectangular dashed box marks the extent of the PV-belt, with an arrow inside the box indicating the starting direction; the solid circle marks the physical scale of the E-bubble and its estimated radius (R $\approx$ 0.75$^\circ$). In the right panel, the solid ellipse and arrows mark the projected physical scale and expansion velocity ($\Delta V$ $\approx$ 6\,km\,s$^{-1}$) of the E-bubble.}
    \label{fig.14}
\end{figure}

Nevertheless, no conspicuous population of massive stars is found inside the Region\,E bubble. We therefore speculate that an as-yet-unidentified supernova may have occurred there. Its swept-up shell is clearly visible in the $^{12}$CO integrated intensity maps (Figure~\ref{fig.14}), but the compact remnant (neutron star) remains undetected, presumably because it received a strong natal kick and is now displaced from the bubble center \citep{Verbunt2017}.

To investigate this possibility, we extracted a position–velocity (P–V) belt cutting across the E-bubble with one end pointing toward ASCC\,125 (Figure~\ref{fig.14}). The bubble has an angular radius of approximately $0.75^\circ$, which at a distance of $\sim$800\,pc (Figure~\ref{fig.7}) corresponds to a physical radius of $\sim$10.5\,pc. The velocity difference between its shell and center is $\sim$6\,km\,s$^{-1}$, indicating a relatively low current expansion velocity. Assuming a supernova origin, a simple kinematic estimate yields a dynamical age of $\sim$1–2\,Myr; if deceleration due to radiative cooling in later stages is considered, the true age may be larger \citep{Cioffi1988}.

Combining this with an estimated mean column density of $10^{21}$\,cm$^{-2}$ (Figure~\ref{fig.8}), the shell kinetic energy is approximately $10^{48}$\,erg. This is substantially lower than the canonical supernova explosion energy of $\sim10^{51}$\,erg \citep{Vink2012}, suggesting several possibilities: (i) the remnant is in a late radiative stage, having lost most of its initial energy through radiation and shock heating \citep{Blondin1998}; (ii) the structure could result from cumulative stellar wind feedback, although no suitable massive star system is identified in the bubble region; (iii) the observed expansion velocity may be underestimated due to projection effects. Regardless of the exact origin, the dynamical impact of this bubble is sufficient to have provided a contemporaneous and significant perturbation for the formation of ASCC\,125.

In other words, the apparent “anomaly” of ASCC\,125 does not reflect unusual initial conditions, but rather the outcome of a more violent evolutionary history. It demonstrates that when feedback is sufficiently intense and the parent cloud is strongly perturbed, star formation efficiency can increase dramatically. Analogous situations have been reported in the nearby Vela\,OB2 and Orion complexes \citep{Beaumont2010, Kounkel2018}.

It should be noted that the extraction of Snake\,III member stars in this work relies on Gaia\,DR3 5-D phase-space filtering, supplemented by quality criteria including $G < 18$\,mag and RUWE $<$ 1.4.
Because a substantial fraction of the stellar complex remains deeply embedded within the clouds of the Cep\,OB3 complex ($A_{\text{V}}$ can reach 5--10\,mag), both MWISP $^{12}$CO emission and 3-D extinction maps reveal pronounced dust gradients and banded extinction structures throughout the region.
High extinction not only drives low-mass stars ($<$ 1\,M$_\odot$) closer to the optical detection limit, but also causes the FoF algorithm in dense cloud cores to preferentially retain bright sources because of infrared blending and source confusion. 
This introduces a selective bias in which the low-mass end of the mass function(MF) is “missed” while the high-mass end is “over-sampled”.
Moreover, many YSOs remain embedded with protoplanetary disks and envelopes, whose intrinsic color offsets can further distort CMD fits and systematically bias mass estimates toward higher values.
Consequently, the slope and turnover of the mass function derived from our present member star sample cannot be directly compared with a universal IMF. 
In follow-up work, we will refine the binary-fraction($f_b$) and mass function(MF) recalibration to provide a clearer portrait of the co-evolution of stars and molecular clouds.

\section{Conclusions}\label{sec:5}

This study focuses on a young stellar filament in the Milky Way known as Stellar “Snake\,III” and its ambient molecular-cloud environment.
The structure contains 5683 member stars with a median age of 7.6\,Myr and spans about $300\times500\times175$~pc$^3$.
We aim to investigate the co-evolution of stars and molecular clouds, with particular emphasis on how cloud structure and stellar feedback jointly regulate star formation progress.

We integrate astrometric data from \textit{Gaia} DR3, CO molecular cloud observations from the MWISP project, and multi-band spectroscopic data from LAMOST.
The member stars of Stellar Snake\,III were isolated by 5-D phase-space filtering (\textit{l}, \textit{b}, $\pi$, $\mu_{l^*}$, $\mu_b$), while the distance-velocity information for the molecular gas allowed foreground and background cloud components to be clearly disentangled, revealing a correlation between stellar age and gas density.
In addition, excitation-temperature and centroid-velocity of the $^{12}$CO molecular clouds, and H$_\alpha$ emission reveal the dynamical coupling between stars and molecular clouds.

The key findings of this work are summarized below:

1. The Nativism and Physical Homogeneity of Snake\,III: 
The astrometric data from \textit{Gaia}\,DR3 and the molecular cloud observations from MWISP collectively support the characteristics of Snake\,III as a native and physically homogeneous structure. 
The high spatial and kinematic consistency between its member stars and the parent molecular cloud indicate that Snake\,III is a native structure with a common origin and evolutionary history.

2. Correlation between stellar age and gas density: 
The overall density of molecular clouds increases with galactic longitude.
The ages of open clusters formed in cavities near molecular clouds in different density regions exhibit a general trend of higher initial cloud density corresponding to older clusters (except ASCC\,125).
Meanwhile, high-density gas environments in existing molecular cloud structures are more prone to give birth to young field stars.
This indicates that the density of molecular clouds has a significant impact on the star formation process.

3. Role of stellar feedback: 
After forming, stars consume part of the gas and then heat and compress the remainder through ionizing radiation and winds, producing shells and bubbles. 
This feedback is clearly evident in all four regions; 
as the feedback intensifies, it will suppress or promote the formation of young stars in the residual gas, demonstrating that feedback regulates both the gas-density distribution and the subsequent star-formation progress.
This feedback process depends not only on the primordial gas density but also on the finely structured distribution of the remaining material.

4. Peculiarity of ASCC\,125: 
Despite being embedded in the highest-density gas environment currently observed, ASCC\,125 exhibits an extremely young age (4.4\,Myr), seemingly breaking the above trend. 
Our analysis shows that, although it sits in the densest part of the cloud, its formation was delayed, but violently accelerated by the combined action of stellar winds from the adjacent cluster UBC\,178 (8.3\,Myr) and a possible supernova blast.
This powerful feedback compressed the surrounding clouds, thereby producing a second-generation stellar system. 

We propose a unified framework in which “the initial and present-day density distribution of a molecular cloud and stellar feedback jointly govern star-formation progress”.

Snake\,III thus constitutes a natural laboratory (7.6\,Myr old and spanning about 300 $\times$ 500 $\times$ 175\,pc$^3$) within which the two competing agents, “cloud-density regulation” and “stellar feedback modulation”, can be tested simultaneously in a single coherent system as drivers of star formation progress.
Future synergy between higher-resolution interferometric data and multi-band observations (e.g., JWST infrared photometry) will enable a recalibration of the binary fraction ($f_b$) and mass function (MF), allowing us to quantify star formation efficiencies (SFE) and feedback histories across individual regions, and thereby place tighter constraints on the co-evolution of stars and molecular clouds.

{\it Acknowledgements.}
The authors thank Feng Wang, Min Fang, Xue-Peng Chen, Ji Yang, and Zhen-Yu Wu for the helpful discussions. H.J.T. thanks the support from the NSFC grant (No. 12373033) and the Key Project of Zhejiang Provincial Natural Science Foundation (No. ZCLZ25A0301). C.W. thanks the support from the NSFC grant (No. 12563004).

\bibliographystyle{aasjournal}
\bibliography{main}
\appendix
The full set of fundamental parameters for the 5683 member stars of Snake\,III (see Table~\ref{table:2} for column descriptions) will be published as a machine-readable table in the online edition of the journal.
\begin{table}[htbp]
    \centering
\caption{The Columns of Snake\,III Catalog}
\label{table:2}
    \begin{tabular}{>{\raggedright\arraybackslash}p{0.02\linewidth}>{\raggedright\arraybackslash}p{0.23\linewidth}>{\raggedright\arraybackslash}p{0.1\linewidth}>{\raggedright\arraybackslash}p{0.6\linewidth}}
        \toprule\toprule
         &  Column&  Unit& Description\\
         \midrule
         1&  Cluster&  $\cdots$& Name of the open cluster to which the star belongs\\
         2&  source\_id&  $\cdots$& Source\_id from $Gaia$$\,$DR3\\
        3&  l& degree&The Galactic Longitude of the celestial object.\\
        4&  b& degree&The Galactic Latitude of the celestial object.\\
        5&  pml\_lsr& mas yr$^{-1}$&The proper motion in Galactic Longitude ($l$) corrected to the Local Standard of Rest (LSR)\\
        6&  pmb\_lsr& mas yr$^{-1}$&The proper motion in Galactic Latitude ($b$) corrected to the LSR\\ 
        7&  X& pc&The X-coordinate in the Galactic coordinate system aligned with the direction from the Sun to the Galactic center\\
        8&  Y& pc&The Y-coordinate in the Galactic coordinate system aligned with the direction of the Galactic rotation\\
        9&  Z& pc&The Z-coordinate in the Galactic coordinate system aligned perpendicular to the Galactic plane, with positive values
        above the plane and negative values below\\
        10&  V\_l& km s$^{-1}$&The tangential velocity component in the direction of Galactic Longitude ($l$)\\
        11&  V\_b& km s$^{-1}$&The tangential velocity component in the direction of Galactic Latitude ($b$)\\ 
        12&  radial\_velocity& km s$^{-1}$&The radial velocity from $Gaia$$\,$DR3$^a$\\
        13&  radial\_velocity\_error& km s$^{-1}$&Error of the radial velocity from $Gaia$$\,$DR3$^a$\\
        14&  rv& km s$^{-1}$&The radial velocity from LAMOST$\,$DR12 LRS$^b$\\
        15&  rv\_err& km s$^{-1}$&Error of the radial velocity from LAMOST$\,$DR12 LRS$^b$\\
        16&  rv\_br1& km s$^{-1}$&The radial velocity from LAMOST$\,$DR12 MRS$^b$\\
        17&  rv\_br1\_err& km s$^{-1}$&Error of the radial velocity from LAMOST$\,$DR12 MRS$^b$\\
        18&  VHELIO\_AVG& km s$^{-1}$&The radial velocity from APOGEE$\,$DR14$^c$\\
        19&  VERR& km s$^{-1}$&Error of the radial velocity from APOGEE$\,$DR14$^c$\\
        20&  pms& $\cdots$&The $pms$ score fit by Sagitta$^d$\\
        21&  age&dex&The log Age fit by Sagitta$^d$\\
        \bottomrule
    \end{tabular}
    \footnotesize
    \begin{flushleft}
    \textit{Note:} $^a$ \citet{Vallenari2023}; $^b$ \citet{Cui2012}; $^c$ \citet{Majewski2017}; $^d$ A trained neural network model \citep[Sagitta;][]{McBride2021} to systematically identify pre-main-sequence (PMS) stars among Snake\,III and to estimate their ages.
    \end{flushleft}
    
\end{table}
\end{document}